\documentclass[11pt,a4paper,fleqn,usenatbib,useAMS]{mnras}
\usepackage[T1]{fontenc}
\usepackage{ae,aecompl}
\usepackage{mathptmx}

\usepackage{graphicx}	
\usepackage{amsmath}	
\usepackage{amssymb}	
\usepackage{multicol}
\usepackage{pdflscape}	
\usepackage{placeins}
\usepackage{float}
\usepackage{xcolor}
\usepackage{enumitem}
\usepackage{ulem}
\usepackage{soul}
\usepackage{hyperref}
\usepackage{url}
\usepackage{breqn}

\title[Short title, max. 45 characters]{Multi-wavelength study of NGC 1365: The obscured active nucleus and off-nuclear compact X-ray sources}

\author[Subhashree Swain et al.]{
Subhashree Swain,$^{1}$
Gulab Chand Dewangan,$^{2}$
P. Shalima$^{3}$,
Prakash Tripathi,$^{2}$
K.V.P. Latha$^{1}$
\\
$^{1}$ \textit{Pondicherry University, Puducherry, 605014, India}\\
$^{2}$ \textit{Inter University Centre for Astronomy and Astrophysics, Pune, India, 411007} \\
$^{3}$ \textit{Manipal Centre for Natural Sciences, Centre of Excellence, Manipal Academy of Higher Education, Manipal, Karnataka 576104, India}\\
}

\pubyear{2022}

\begin{document}
\label{firstpage}
\pagerange{\pageref{firstpage}--\pageref{lastpage}}
\maketitle

\begin{abstract}
We present a multi-wavelength study of the active nucleus and the off-nuclear X-ray sources in the nearby spiral galaxy, NGC~1365 using three simultaneous UV/X-ray observations by \textit{AstroSat} over a two months period and archival IR observations performed with \textit{Spitzer} and \textit{Herschel}. 
Utilising the data from the Soft X-ray Telescope (SXT) on-board  \textit{AstroSat}, we find spectral variability mainly caused by the variation in the X-ray column density, (N$_H$ $\sim$ 10$^{22}$ - 10$^{23}$ \rm{cm}$^{-2}$). With the accurate spatial resolution of the UVIT onboard \textit{AstroSat}, we separate the intrinsic AGN flux from the host galaxy emission and then correct for the Galactic and the internal reddening. We detect no significant variation in the NUV emission over the observation period. The AGN in FUV band is undetectable due to heavy intrinsic extinction. Further, the multi-wavelength IR/UV/X-ray AGN SED reveals that the AGN is in a low luminosity phase with accretion rate $\sim$ 0.01 L$_{Edd}$. The steady UV emission and strong X-ray absorption variability suggest that the obscuring clouds are likely compact and affect the compact X-ray source only and do not possibly cover the extended UV emitting region. In addition, the UVIT is able to resolve two bright spots at a radius of 7$\arcsec$ ($\sim$ 6.3 Kpc) from the central nucleus in the South-West (SW) direction. In the UVIT image of the entire galaxy, we identify UV counterparts to four {\it Chandra} identified bright X-ray sources. One well-known ultra-luminous X-ray source (ULX) NGC~1365 X2 is identified with its UV counterpart at 86$\arcsec$ from the nucleus in the north-east (NE) direction from the active nucleus.
\end{abstract}




\section{Introduction}

Unification scheme of active galactic nucleus (AGN) relies on the existence of an anisotropic torus in the nuclear structure, which leads to a diversity of AGN properties at different wavelength bands. 
The torus is the key to unlocking information about the immensely energetic AGN, as the fraction of obscuration plays an important role in determining the type of AGN.
The torus emits in the infra-red (IR) band by absorbing UV photons from the accretion disk \citep{Netzer2015, almeida2011testing}.
For many decades already, models from X-rays to IR spectra and broadband SEDs have been the only means to derive clues on the dust geometry, composition, and its distribution.
Furthermore, signatures of reprocessed emissions by
the torus in the X-ray band arises primarily from the interaction of X-ray photons from the corona or accretion disk with the surrounding gas \citep{1994MNRAS.267..743G, 1994ApJ...420L..57K}. Hence the IR emission from active galaxies is due to a combination of reprocessed higher energy emission and thermal emission from star formation. Thus, it is crucial to separate intrinsic emission from the AGN. 
The torus and the circumnuclear gas can also induce X-ray variability
via variable obscuration. Depending on the absorber's location, the characteristic timescale of the latter can be from months to years
if the absorber is part of a distant torus \citep{rissalti2002, miniutti2014}, or from days to weeks if the absorber
is located closer in the so-called broad-line region, e.g. NGC
4388 \citep{Elvis_2004}, NGC 1365 \citep{Risaltti2009}, NGC
4151 \citep{puccetti_2007}, NGC 7582 \citep{bianchi_2009},
 and Fairall 51 \citep{svoboda_2015}. Generally, type 1 AGN such as Seyfert 1s exhibit intrinsic soft X-ray variation as the central engines can be viewed directly, but observing soft X-ray variability in Seyfert 2s is a great challenge as the central engines are obscured by the torus. Although it is well established that a number of Seyfert 2s are variable, it is unknown whether the same kind of variation is common for all the nuclei or, more importantly, what drives these variations.
 The observed variations in the Seyfert 2s may be related to the absorbing material that crosses observer's line of sight (LOS) \citep{Risaliti_2002, Risaliti_2010} and/or can be intrinsic to the sources \citep{evans_2005, braito_2013}. A few Seyfert 2s also show changes from being reflection-dominated to transmission-dominated objects, hence named as changing-look objects \citep{Guainazzi_2002},  \citep{Risaliti_2010}. These changing look objects are mainly associated with dust clouds from BLR or torus along the LOS. The composition and distribution of dust inside the torus are one of the hot debated topics (\cite{1992ApJ...401...99P}; \cite{nenkova2008}; \cite{Feltre2012}; \cite{subhashree2021}).
 
We selected a nearby obscured AGN NGC~1365
as it has both nuclear and star formation activities. This galaxy is a member of the Fornax cluster with a distance of $\sim$ 18.6~Mpc $\pm$ 0.6 \citep{Madore1999,  Silbermann_1999, springob2009, polshaw2015, jang2018}. NGC~1365 has a Seyfert nucleus, a prominent bar, and is an archetype barred spiral galaxy (SBb(s)I;\cite{1981RSA...C...0000S}). Dark dust lanes run across the nuclear region and partially obscure the nucleus. A thorough review of NGC~1365  is given in \cite{1999A&ARv...9..221L}. It hosts vigorous star formation in the circumnuclear region, and a variable obscured Seyfert 1.8 nucleus \citep{Risaltti2009}.
Actually, this AGN is classified as Seyfert 1, 1.5, 1.8 or 2 in the literature (e.g., \cite{veron1980}; \cite{turner1993}; \cite{mailino1995}; \cite{risaliti2007}; \cite{thomas2017}) due to variation in the X-ray flux and  the column density of the absorbers along the line of sight \citep{gracia2015}. Apart from the scenario of dust clouds in BLR, disk winds have recently been used as a mechanism to explain the variable column density \citep{connolly_2014, braito_2013, Liu_2021, Mondal2022}. In this work, we study X-ray/UV variation and also investigate the off-nuclear X-ray sources in NGC~1365 as Ultraluminous X-ray sources (ULXs) are associated with star formation.
\\

 Here we use simultaneous UV/X-ray observations performed with \textit{AstroSat} and IR observations performed with archival IR data acquired with \textit{Spitzer} and \textit{Herschel}. This paper is organized as follows. Section 2 describes the observations and data reduction. The imaging data analysis and spectral analysis are described in Section 3. The implications of our results are discussed in section 5, and the conclusions in section 6.

\section{Observations and data reduction}
 The \textit{AstroSat} (\cite{singh2014}: \cite{AGRAWAL20062989}) carries four co-aligned instruments such as Ultra-Violet imaging Telescope (UVIT), Soft X-ray Telescope (SXT), Large Area X-ray Proportional Counter (LAXPC) and Cadmium-Zinc-Telluride Imager (CZTI). As the source is too faint, it is not detected in LAXPC or CZTI. Thus, we used the data acquired with the UVIT and SXT only. 
 Three sets of observations of NGC 1365 were performed during November - December 2016 for NGC 1365. 
 The details of these observations are listed in Table \ref{tab:tab1}.
We used the data from UVIT and SXT. We also used X-ray data acquired with the {\it Chandra} and IR data from \textit{Herschel} and {\it Spitzer} as listed in Table~\ref{tab:tab1}.
\begin{table}
	\centering
	\caption{Observation log of the data used in our spectral modeling.}
	\label{tab:tab1}
	\tiny
	\begin{tabular}{lcccr} 
		\hline
		Observatory & OBSID & Date& Exposure time & FWHM \\
		\hline
	
		\textit{AstroSat}/SXT & 9000000776 & 2016 Nov 07& 19.9 ks & 0.2$\arcmin$\\
		\textit{AstroSat}/SXT & 9000000802 &2016 Nov 17&  17.6 ks\\
		\textit{AstroSat}/SXT & 9000000934 &2016 Dec 27& 22.6 ks\\
		\hline
		\textit{AstroSat}/UVIT/FUV & 9000000776 &2016 Nov 07&  7.9 ks & 1.3 $-$ 1.5$\arcsec$\\
		\textit{AstroSat}/UVIT/FUV & 9000000802 &2016 Nov 17&  12.1 ks\\
		\textit{AstroSat}/UVIT/FUV & 9000000934 &2016 Dec 27&  6.7 ks\\
		\hline
		\textit{AstroSat}/UVIT/NUV & 9000000776 &2016 Nov 07&  9.9 ks &1 $-$ 1.4$\arcsec$\\
		\textit{AstroSat}/UVIT/NUV & 9000000802 &2016 Nov 17& 11.8 ks\\
		\textit{AstroSat}/UVIT/NUV & 9000000934 &2016 Dec 27& 6.4 ks\\
		\hline	\textit{Chandra}/ACIS & 6868 &2006 Apr. 17& 14.6 ks & 0.5$\arcsec$\\
		\textit{Chandra}/ACIS & 6869 &2006 Apr. 20& 15.5 ks\\
		\textit{Chandra}/ACIS & 6870 &2006 Apr. 23& 14.6 ks\\
		\textit{Chandra}/ACIS & 6871 &2006 Apr. 10 &13.4 ks\\
		\textit{Chandra}/ACIS & 6872 &2006 Apr. 12& 14.6 ks\\
		\textit{Chandra}/ACIS & 6873 &2006 Apr. 15& 14.6 ks\\
		\hline
		\textit{Spitzer}/MIPS & 631 &2004 Dec. 24& 0.52 ks & 5.9$\arcsec$\\
		\hline
		\textit{Herschel}/PACS (70 $\mu$\rm{m}) & 1342183551 &2009 Sept. 09& 1.3 ks & 5.6$\arcsec$\\
		\textit{Herschel}/PACS (100 $\mu$\rm{m}) & 1342183552 &2009 Sept. 09&1.3 ks & 6.8$\arcsec$\\
		\textit{Herschel}/PACS (160 $\mu$\rm{m}) & 1342183553 &2011 June 11& 1.3 ks& 11.3$\arcsec$\\
		
		\hline
	\end{tabular}
\end{table}

\subsection{The \textit{SXT} data}
The SXT \citep{2017JApA...38...29S} is India's first X-ray telescope based on the principle of grazing incidence. 
The SXT has low resolution for imaging (full width at half maximum, FWHM $\sim$ 2$\arcmin$) with a circular field of $\sim$~40$\arcmin$ diameter.
X-rays in the
energy band of 0.3 $-$ 8.0 keV are focused on a cooled charge coupled device, thus providing medium resolution
X-ray spectroscopy of cosmic X-ray sources of various types.  
 We used processed and cleaned level2 event files directly from the \textit{AstroSat} archive. We merged the  event files of each observation ID using SXTMerger.jl tool\footnote{\url{https://github.com/gulabd/SXTMerger.jl}}, also available at the SXT/POC website\footnote{\url{https://www.tifr.res.in/~astrosat_sxt/dataanalysis.html}}. Finally, we used the XSELECT tool, available within the  HEASOFT (version 6.25) software, to extract the source spectrum from the merged event list. 
According to most recent studies of the SXT point spread function (PSF) of size 2$\arcmin$ and half power diameter of 10$\arcmin$, a circle of 15$\arcmin$ radii centered at the source position is sufficient to get 95\% of the total photons from the target. The large extraction is required due to the extended wings of the PSF. It is also considered for the lowest possible systematic. We may lose photons using a small region ($\sim$ 5$\arcmin$ radius), resulting in a low signal to noise ratio. In addition, the response and background files are made for 15$\arcmin$ by the {\it AstroSat}-SXT team members. Thus we extracted the source spectrum from the merged event list using a circular region of 15$\arcmin$ radius centered on the source. As SXT CCD camera has no source free regions, we used blank sky background spectrum 
SkyBkg\_comb\_EL3p5\_Cl\_Rd16p0\_v01.pha provided by the SXT instrument team. We also used spectral redistribution matrix file (sxt\_pc\_mat\_g0to12.rmf). We used an updated ancillary response file generated by the team based on the original file 
(sxt\_pc\_excl00\_v04\_20190608\_corr1kev\_sharp\_chng\_crab2\_
15arcmin\_d2m\_1\_12kev.arf).
Finally, we grouped the spectral data to a minimum of 30 counts per bin. 
In addition to SXT observations, we also utilized {\it Chandra} archival data (0.2$-$10 keV) from the
Advanced CCD Imaging Spectrometer spectroscopy array (ACIS-S; \cite{gamire2003}) acquired with the observations performed in 2006 April, with a total exposure time of ~84 ks. We used the level2 data directly for the six observations from Table \ref{tab:tab1} for our study. We used the {\it Chandra} data for imaging study in our case.
\subsection{The \textit{UVIT} data}
We performed a deep NUV and FUV imaging observations of NGC 1365 using the UVIT 
(UVIT;\cite{Tandon2017, Tandon2020}) simultaneously with the SXT observations. The UVIT consists of two co-aligned telescopes, one for FUV (1300 $-$ 1800 \AA) and
another for both NUV (2000 $-$ 3000 \AA) and visible (VIS) channel (3200 $-$ 5500 \AA). The UVIT is capable of observing sources simultaneously
in all three available bands. The visible channel
is mainly used to track the drift pattern of the satellite. Each of
the pointing telescopes has a 28$\arcmin$ circular field of view with an
angular resolution, FWHM 1$-$1.5$\arcsec$.
Both UV channels of UVIT are equipped
with multiple photometric filters, which provide a unique
imaging capability in the ultra-violet bands. NGC 1365 was imaged in FUV/F169M ($\lambda_{mean}$ = 1608 \AA, $\bigtriangleup\lambda$ = 290 \AA) and NUV/N279N ($\lambda_{mean}$ = 2792 \AA, $\bigtriangleup\lambda$ = 90 \AA) filters. We obtained the UVIT level1 data from \textit{AstroSat} archive and processed them  using UVIT pipeline CCDLAB \citep{Postma_2017}. CCDLAB takes care of drift correction and alignment of data from different orbits. We generated clean images for each observation and merged them. We finally produced  deep
images of 28.2 ks exposure time in NUV and 26.7 ks in FUV bands.
\par
\begin{figure*}
		\includegraphics[width=15cm]{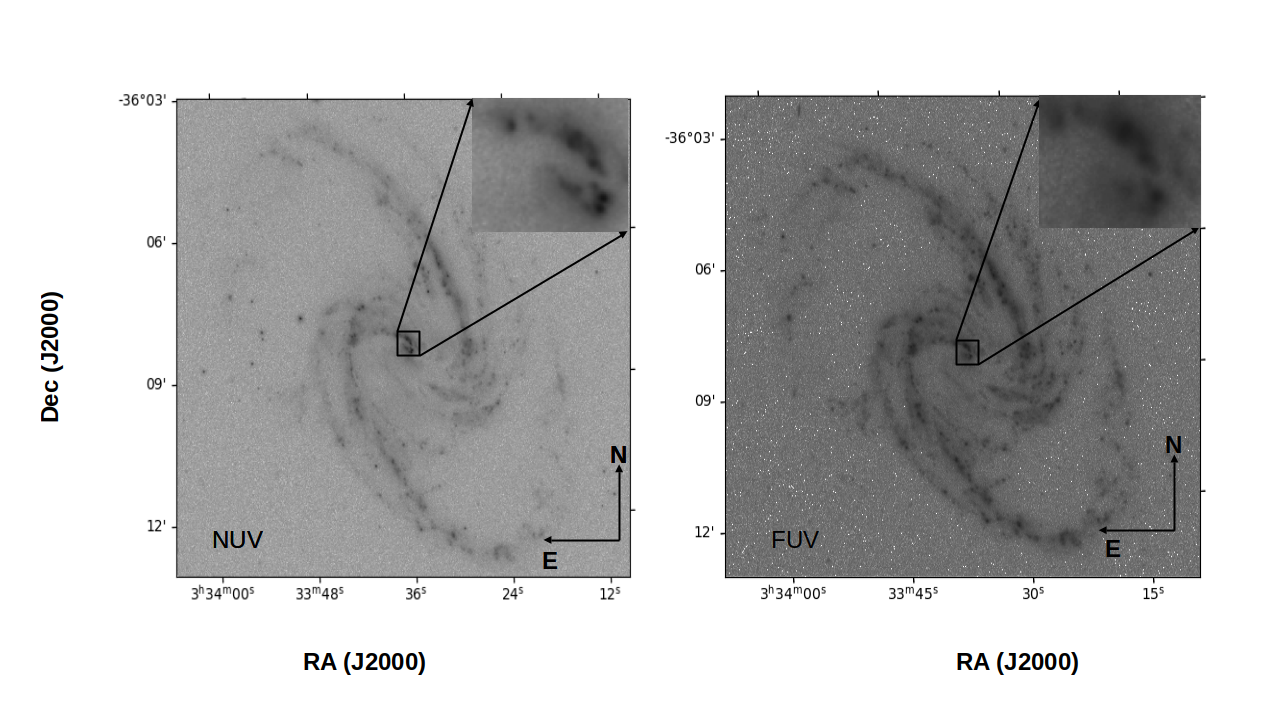}
    \caption{The UVIT NUV//N279N (left) and FUV/F169M (right) image of NGC~1365  with size 10$\arcmin$ $\times$ 10$\arcmin$ with a zoomed-in central portion of size 40$\arcsec$ $\times$ 45$\arcsec$. The active nucleus is not seen in the NUV and the FUV bands due to heavy extinction by the dust lane in the host galaxy passing through the central region.}
    \label{fig:fig1}
\end{figure*}

We performed astrometry on UVIT images  by converting the image coordinates of sources to the WCS coordinates by using IRAF 2.16 `imcoords' task. 
Astrometry on each UVIT image was done by cross-matching 5 field stars in the \textit{SWIFT} Ultraviolet Optical Telescope (UVOT) image as reference image which has $\sim$ 2$\arcsec$ resolution. 
 The FWHM of reference stars in the UVIT images are $\sim$ 1--1.5$\arcsec$. The astrometry corrected UVIT images in NUV and FUV bands are shown in Figure~\ref{fig:fig1}. The UVIT images have
4096$\times$4096 pixels where 1 pixel corresponds
to 0.4$\arcsec$, which is equivalent to  $\sim$ 36 pc at the distance
of NGC 1365. The final images in both the bands suggest that the central region of the AGN in the FUV is more extinct than that in the NUV band.
In our study, we have analyzed the data from 6$\arcsec$ aperture of the central nucleus in order to reduce the host galaxy contribution. We do not see a clear point source at the center of NGC 1365 in both the NUV and FUV images. This is likely due to the obscuration by the torus or a dust lane passing through the nucleus. 
We also constructed a composite color image by combining FUV, NUV and Chandra/ACIS-S X-ray images, the color images, shown in Figure \ref{fig:fig3}, helps in identifying the location of the nucleus.

\subsection{The IR data}
\begin{figure}
	\includegraphics[width=\columnwidth, height=6cm]{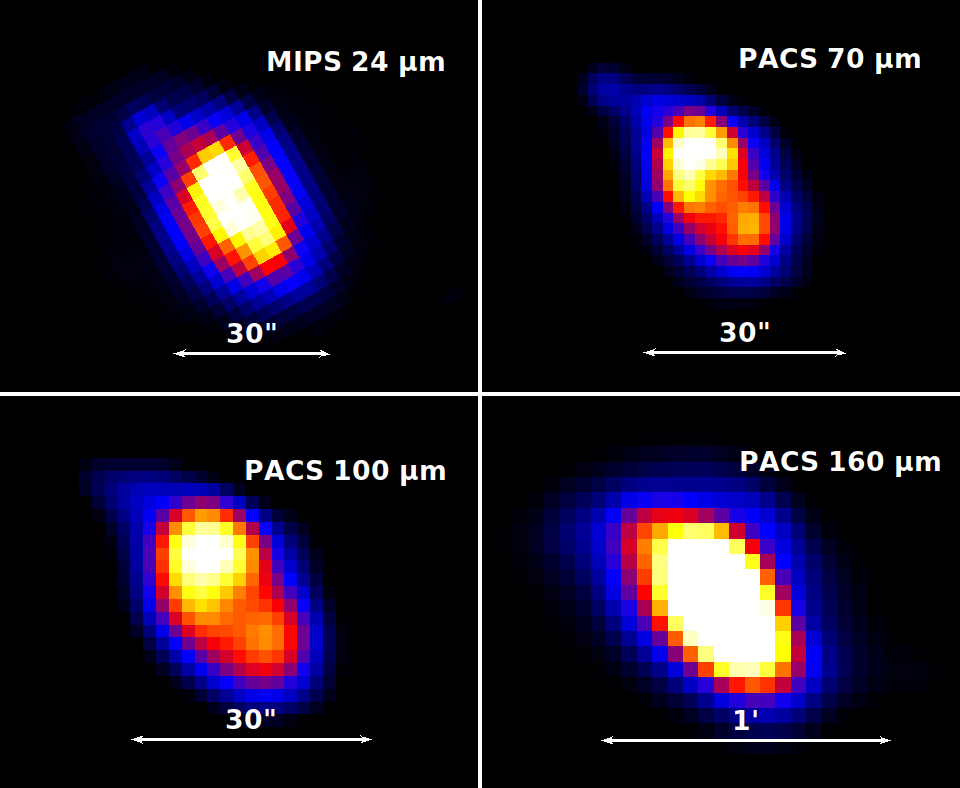}
    \caption{ The IR image of NGC 1365 (1.5$\arcmin$ $\times$ 1$\arcmin$) in different bands. The cross mark (black color) represents the {\it Chandra} X-ray position of the AGN.}
    \label{fig:fig4}
\end{figure}

To determine the SED spanning mid-IR to far-IR energies, we have used the data acquired by \textit{Spitzer} and \textit{Herschel}. 
We used archival Spitzer data taken with the Multi-band Imaging Photometer for \textit{Spitzer}
(MIPS; \cite{Rieke2004}). Although 24 $\mu$\rm{m} has a poor angular resolution, the AGN is still sufficiently bright, and hence can be distinguished.
We have also utilized
far-infrared (FIR) data from Herschel PACS (70 and 160 $\mu$\rm{m}
filters) and SPIRE (250 $\mu$\rm{m}). The data are part of the guaranteed time program entitled
`Herschel imaging photometry of nearby Seyfert galaxies: testing
the coexistence of AGN and starburst activity and the nature of the
dusty torus' (PI: M. Sánchez-Portal). \cite{hererroetal2012} compared the morphologies of the PACS 70 $\mu$\rm{m} and the MIPS 24 $\mu$\rm{m} images as they have similar angular resolutions. We use these two images primarily for PSF to extract the flux of AGN. The PSF of the central source in the image of 24 $\mu$\rm{m} is about 7.48$\arcsec$. The Moffat PSF of the source in \textit{Herschel}/PACS 70 $\mu$\rm{m} is 6.21$\arcsec$. We have used central circular region of 6" radius for the study of AGN, which is comparable to the size of the PSF. All the IR images are shown in Figure \ref{fig:fig4}. 
The stacked pictures of the NUV, FUV and IR bands are displayed in Figure \ref{fig:fig5}.
 Here the AGN nucleus shown in the Figures is identified from $Chandra$ X-ray image by \cite{Wang_2009}. 

\section{Analysis and results}
\begin{figure*}
	\includegraphics[width=18cm]{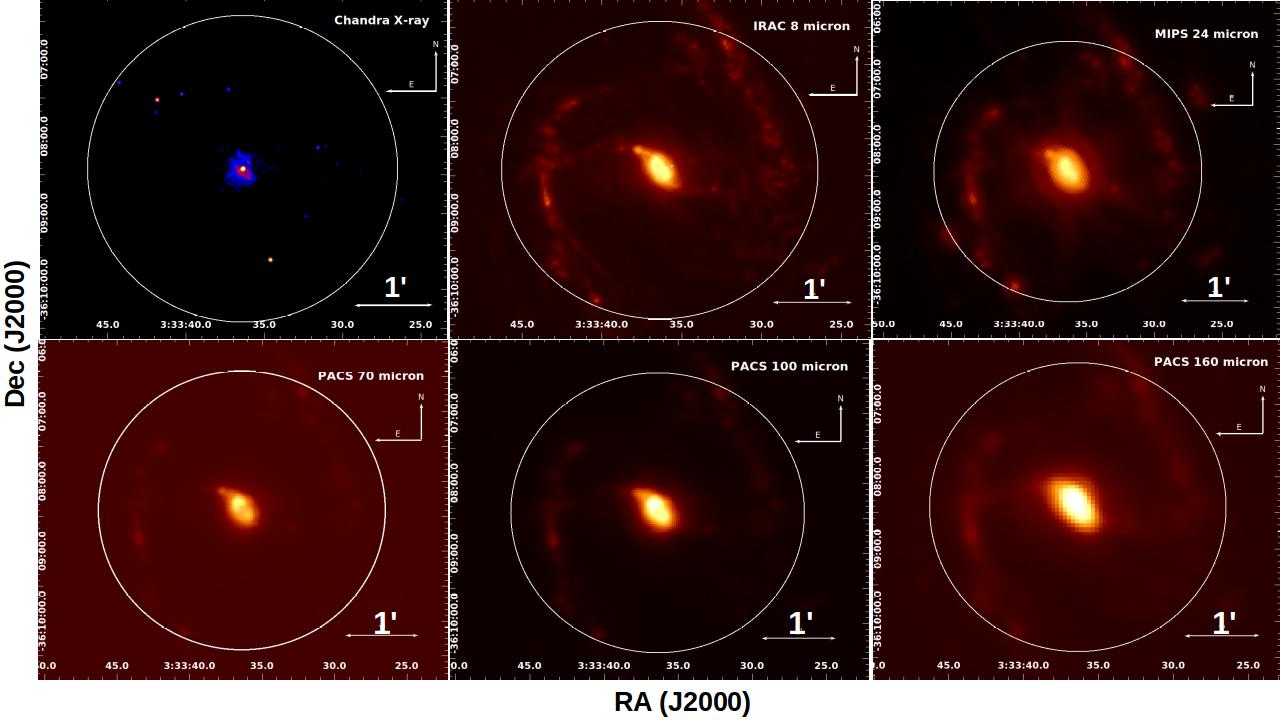}
    \caption{The smoothed images of NGC 1365 galaxy in the X ray band(0.3 $-$ 8 keV), IRAC 8 $\mu$\rm{m} band, MIPS 24 $\mu$\rm{m} band, PACS 70 $\mu$\rm{m} band, PACS 100 $\mu$\rm{m} band, PACS 160 $\mu$\rm{m} band containing the central bright nucleus respectively for an easy comparison between the images of different spatial resolution. The angular resolutions of the Chandra image is 0.5$\arcsec$ and IRAC image is 2$\arcsec$ (FWHM) and that of the MIPS 24 $\mu$\rm{m}, PACS 70 $\mu$\rm{m} and PACS 100 $\mu$\rm{m} is $\sim$ 6$\arcsec$. The X-axis is RA and Y-axis is Dec. North is up and the east is to the left in each panel.}
    \label{fig:fig50}
\end{figure*}

\begin{figure}
	\includegraphics[width=\columnwidth]{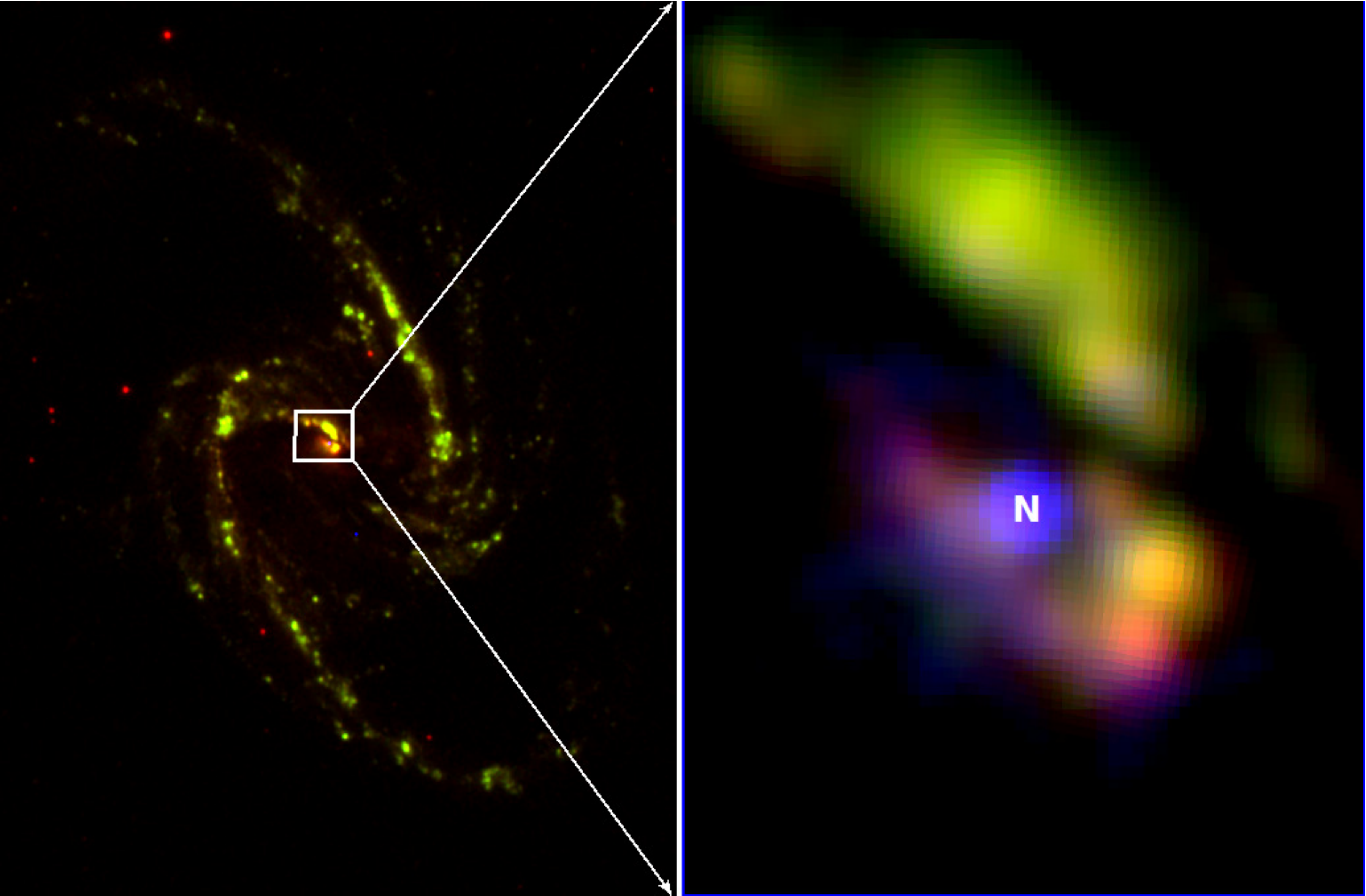}
    \caption{Color composite image of NGC 1365 constructed from stacked {\it Chandra}/ACIS-S image in the X-rays (blue), UVIT/NUV (red) and FUV (green) bands. The rectangular region presents 0.3$\arcmin$ $\times$ 0.4$\arcmin$ inner region that includes the active nucleus marked with "N". For both images, orientation is north up, east to the left.}
    \label{fig:fig3}
\end{figure}

In this section, we present the results of the simultaneous X-ray/UV observations. We used the images in the Chandra X-ray (0.3--8 keV), NUV (N279N filter), FUV (F169M filter), mid-IR (MIPS 24 $\mu$\rm{m}), far-IR (Pacs 160 $\mu$\rm{m}) bands and constructed composite 3-color images with different combination of bands. Figure \ref{fig:fig50} depicts the distinct images that have been utilized to create composite images for the case study. 
Figs. \ref{fig:fig3} and \ref{fig:fig5} show the 3-color combination of X-ray, NUV, FUV and
\textit{Spitzer} 24 $\mu$\rm{m} images of NGC 1365.

\subsection{X-ray-UV-IR morphology}
\label{sec3.1}
\begin{figure*}
	\includegraphics[width=5.5cm,height=5cm]{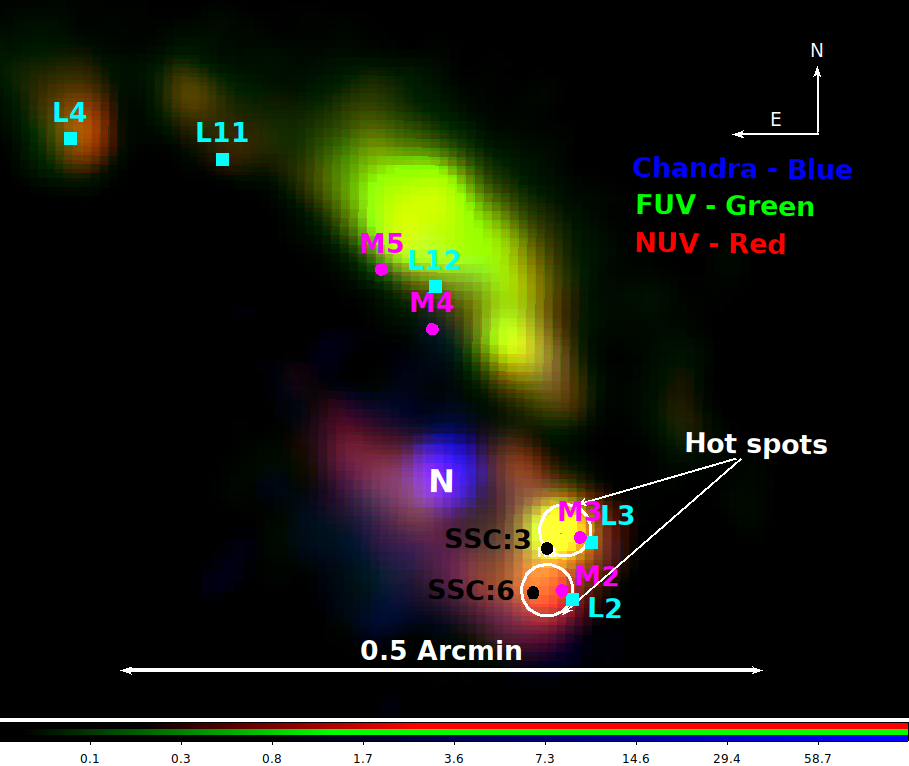}
	\includegraphics[width=5.5cm,height=5cm]{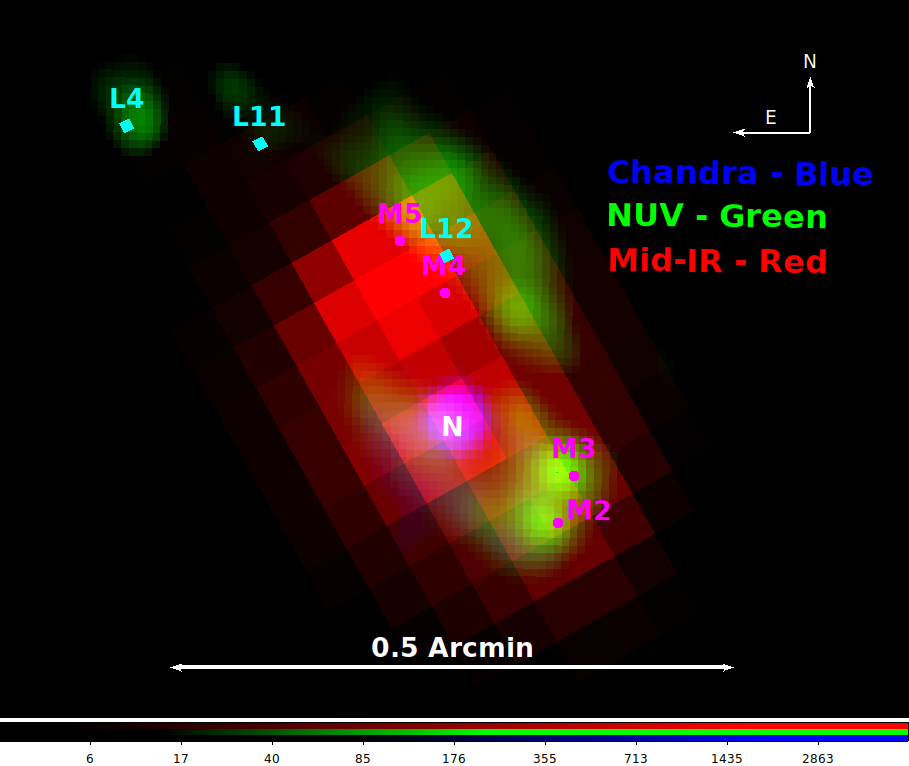}
	\includegraphics[width=5.5cm,height=5cm]{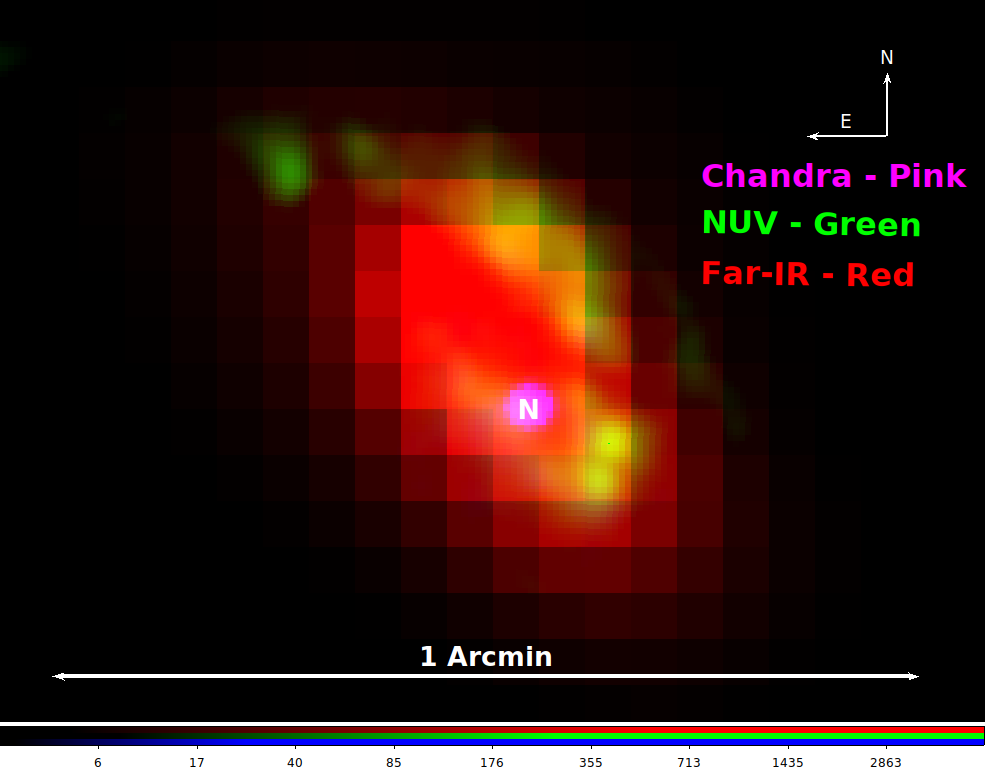}
    \caption {First Panel: The smoothed color image of stacked X ray (0.3 $-$ 8 keV), NUV and FUV of the inner 36$\arcsec$ $\times$ 36$\arcsec$ region showing the location of the AGN (`N'), two bright hot spots (white circle symbols) from UVIT instrument, the mid-IR clusters M2...M5 (filled magenta circle symbols) of \protect\cite{Galliano2005}, the H$\alpha$ hot spots L2, L3, L4, L11, L12 (filled cyan squares) of \protect\cite{alloin1981} and \protect\cite{kristen1997} and super star clusters (SSC) (filled black circles symbols) from \protect\cite{kristen1997}. Middle panel: The stacked image in X-ray, NUV and MIPS 24 $\mu$\rm{m} of the inner 72$\arcsec$ $\times$ 72$\arcsec$ region containing AGN. The `N' represents $Chandra$ X ray nucleus, green represents NUV and red color represents MIPS 24 $\mu$\rm{m} image. Last panel: The smoothed color image of stacked X-ray in pink, NUV in green and PACS 160 $\mu$\rm{m} in red with `N' as the nucleus.}
    \label{fig:fig5}
\end{figure*}

Figure \ref{fig:fig50} represents the resolution wise morphology of AGN. {\it Chandra} image has a very good resolution (0.5$\arcsec$). Then the UV image has 1.3$-$1.5$\arcsec$ resolution where the nucleus identified by {\it Chandra} is obscured by the dust lane. The central bright region of NGC 1365 is detected in the IR band, where the {\it Spitzer} and {\it PACS} image has 6$\arcsec$ spatial resolution. The two bright spots detected by UVIT are also obscured by the dust in Fig. \ref{fig:fig50}. Hence the purpose of overlap with images as shown in Figure \ref{fig:fig3} is to infer the multiwavelength scenario of the central nuclear region. The composite images are only used for identifying the AGN and off-nuclear sources in the entire galaxy, not for the precise kinematics.
The composite images are shown in 
Fig. \ref{fig:fig5}. Each panel shows the AGN with a bright nucleus and its circumnuclear regions.  
As seen in Fig. \ref{fig:fig5}, there are two partially extended hot and bright spots located to the south-west of the nucleus at a distance of $\sim$ 7$\arcsec$ (647 pc) from the AGN. The circumnuclear star formation ring is elongated in the NE-SW direction of the nucleus. They are commonly observed in barred spiral galaxies and connected with Inner Lindblad Resonance (ILR) \citep{1999A&ARv...9..221L}. There is a dark dust lane across the middle of the ring \citep{hererroetal2012} which is prominent in Infra-red (IR) as shown in the middle and right panel of Fig. \ref{fig:fig5}, where we show a close-up of the inner 36$\arcsec$ $\times$ 36$\arcsec$ region. 
We marked the positions of the bright (designated as M4, M5) and faint mid-IR (designated as M2 and M3) clusters using the relative positions given by \cite{Galliano2005}. We also indicated the positions of H$\alpha$ sources  from \cite{alloin1981} and \cite{kristen1997}. 
We do not see any offset between $Chandra$ position and the IR position of AGN in the middle panel of Fig \ref{fig:fig5}.

\begin{figure*}
\centering
	\includegraphics[width=17cm,height=6cm]{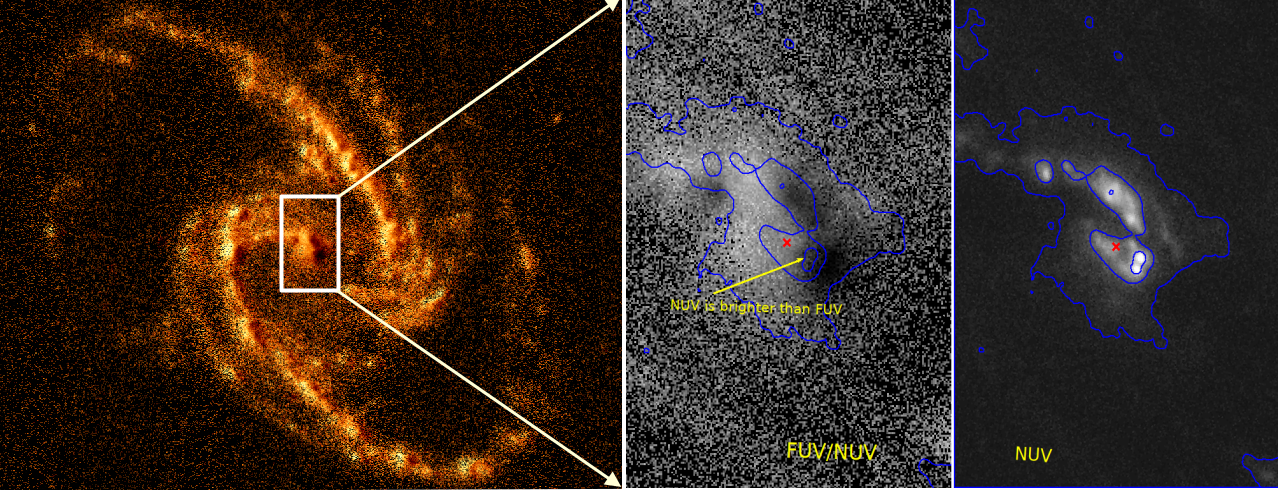}
    \caption{Left panel: The color map (FUV/NUV flux) obtained from UVIT instrument for NGC~1365, Middle panel: The color map for zoom version of the central nucleus with the contour levels, Right panel: The zoom version of the central nucleus with the same contour levels in the NUV band. The red cross on the figure represents the position of the central nucleus. }
    \label{fig:figcm}
\end{figure*}

We created FUV/NUV colour map of the galaxy as presented in Figure \ref{fig:figcm}. We calculated the magnitude using zero point magnitude from \cite{Tandon2017} and found that the extinction and background corrected color index m$_{NUV}$--m$_{FUV}$ is -- 0.92 $\pm$ 0.25 in the centre of NGC1365 for mean of the three observations. It infers that NUV emission is more dominant than FUV.
The overall morphology of the UV emission is as follows.
\begin{enumerate}[label=(\roman*)]
\item The AGN is more extinct in the FUV than NUV (see Fig. \ref{fig:fig1}).
\item Two incomplete cones are extended in NE-SW direction from the position of the nucleus, which is different from soft X-ray emission cone \citep{Wang_2009} and optical cone \citep{lindblad1996} . The gap between the two cones is completely obstructed by IR emission which suggests an active star formation.

\item The compact circumnuclear ring maintains regularity close to the center. The inner ILR and outer ILR in NGC 1365 are $\sim$ 3$\arcsec$ and $\sim$ 27$\arcsec$
 from the nucleus, respectively \citep{hererroetal2012}. The UVIT detected ring is located at $\sim$ 7$\arcsec$ from the nucleus which is outside of the inner ILR. 
\item Two bright spots in the circumnuclear region at a distance of $\sim$7$\arcsec$ from the nucleus are detected in the NUV band, suggesting star forming regions. It is well in agreement with the optical  H$\alpha$ hot spots and MIR regions \citep{hererroetal2012} as shown in Fig. \ref{fig:fig5}. The region contains bright spots which coincide with the individual compact star clusters and mid-IR clusters as shown in the left and middle panels of Fig. \ref{fig:fig5}.
\end{enumerate}

\subsection{The off-nuclear X-ray sources with their UV and IR counterparts and environments}
\label{sec3.2}
\begin{figure}
	\includegraphics[width=\columnwidth]{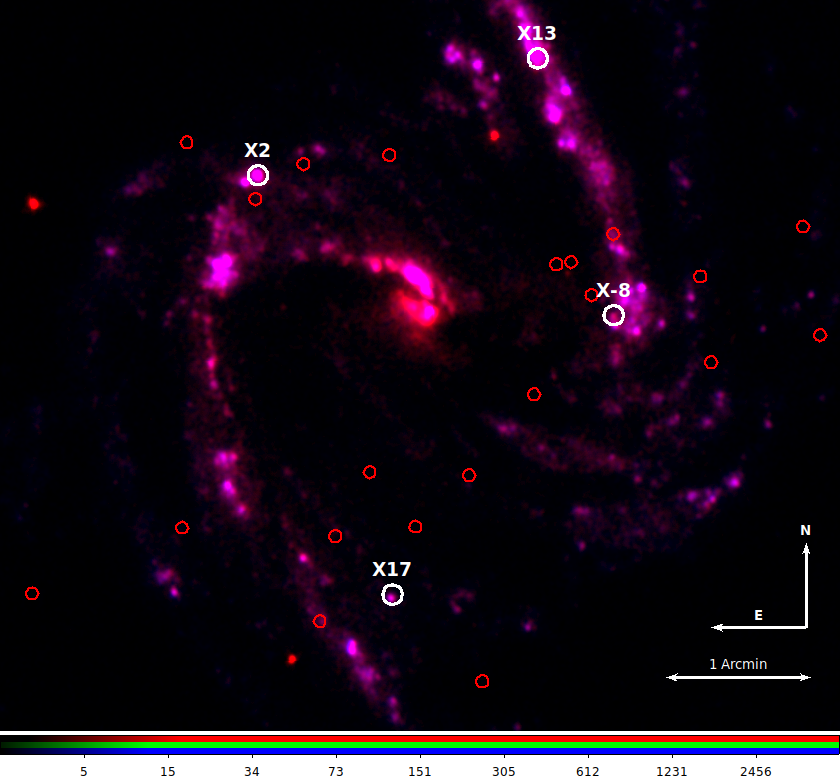}
    \caption{Color (UVIT, NUV \& FUV) image of the inner 6$\arcmin$ $\times$ 6$\arcmin$ ($\sim$ 36 kpc $\times$ 36 kpc) of NGC 1365. The 26 X-ray point sources identified by {\it Chandra} are shown as circles. North is up and the east is to the left. The white circle represents the UV counterparts of the X-ray sources while the red circles represents the X-ray sources that are not detected in the NUV or FUV image.}
    \label{fig:fig10}
\end{figure}

\begin{figure}
	\includegraphics[width=\columnwidth]{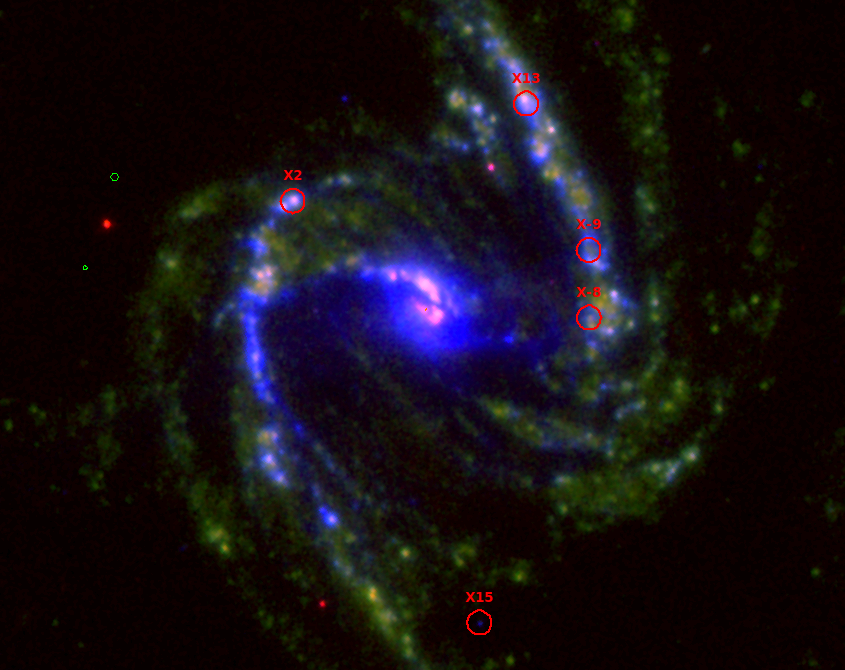}
    \caption{Color (IRAC 8 $\mu$\rm{m}, NUV \& FUV) image of NGC 1365 with X-ray sources in the red circle. Red circles are of radius 6$\arcsec$, which coincides with UV/IR counterparts.}
    \label{fig:fig12}
\end{figure}

NGC 1365 is a very faint source in 
SXT 2016 observations as SXT has less effective area of 90 cm$^2$ at 1.5 keV and medium energy resolution of
80 -- 150 eV in the 0.3 -- 8.0 keV range. It is capable of low resolution imaging with the Point Spread Function (PSF) having a full width half-maximum of $\sim$120". Hence SXT cannot separate out ULX sources well. Thus, we use {\it Chandra} observations to show the off-nuclear sources.
The {\it Chandra} observations showed extended emission with the detection of 23 off-nuclear point sources in 2002 and 3 other sources in 2006 \cite{Strateva_2009} in the central $\sim$ 6$\arcmin$ $\times$ 6$\arcmin$ region. Four X-ray sources can be identified in  UVIT images with  their UV counterparts. (see Figure \ref{fig:fig10}). They are X2, X13, X17 and X8 as named in \cite{Strateva_2009}. 
In order to identify IR counterparts of these X-rays sources, IRAC 8 $\mu$\rm{m} and \textit{Spitzer} 24 $\mu$\rm{m} images are used and overlaid with X-ray sources identified by {\it Chandra}. The IRAC 8 $\mu$\rm{m} image (Figure \ref{fig:fig12}) contains five different infrared counterparts (X2, X8, X9, X13, and X15), while MIPS 24 $\mu$\rm{m} image contains two different infrared counterparts (X2, X13). We listed UV count rate for UV counterparts detected in the UVIT image in Table \ref{tab:tab25}.

\subsubsection{UV counterparts}
\begin{enumerate}[label=(\roman*)]
\item The UV counterpart of one of the X-ray point sources, NGC~1365 X2 is a highly variable X-ray point source in the north tip of the eastern spiral arm of NGC 1365. It is a transient source with intermediate blackhole (IMBH) mass of $<$ 50M$_\odot$ with co-spatial UV emission. It has UV counterparts in UVIT image and optical counterparts from the literature \citep{Strateva_2009}. It is associated with a region containing a bright spot with FWHM $\sim$ 2$\arcsec$ (see
Fig. \ref{fig:fig10}, 86$\arcsec$ south from NGC 1365 nucleus (equivalent
to $\sim$ 8.8 kpc)), which was classified as a HII region by \cite{hodge_1969}. The source X2 is around 3\% and 2\% of the peak luminosity \citep{Strateva_2009} in the NUV band and the FUV band respectively. 
Hence it is highly unlikely that X2 is a background AGN. 
This UV/HII region makes a dense environment for ULX source to have an IMBH.  
The ULX source X2 has more FUV emission than NUV emission as appeared in the UVIT image from Table \ref{tab:tab25}, which is highly unlikely to predict that the ULX might be hosted by globular cluster \cite{Strateva_2009}. In general, stars in globular clusters are expected to be older. However, the strong FUV emission of sources may be responsible for the formation of young stars in globular cluster \cite{sahu2022} or might be associated with young stars clusters as in the case of \cite{poutanen2013}. However, \cite{Strateva_2009} estimated an X-ray-to-optical ratio of --2 $<$ log [X/O] $<$  --1, where we estimated a X-ray-to-UV ratio of --1.52 and --1.7 for NUV and FUV bands respectively. It suggests that UV and optical emission are dominant in the counterpart region which may harbour IMBH. Here, we deferred the details of the investigation.

\item X8 is co-spatial with faint NUV/FUV emission at 82$\arcsec$ from the nucleus along the west direction of the right side spiral arm. 
Our understanding of this source is limited because we don't know whether they are a part of the galaxy or not, and also due to lack of spectral analysis.

\item X13 is located in an active HII zone \citep{royetal1997} with UV extended emission. It is separated by $\sim$ 4$\arcsec$ from its UV counterpart within the circle.  
\item X17 is one of the faint X-ray sources with UV/optical counterparts located at 120$\arcsec$ from the nucleus. It is separated by $\sim$ 2$\arcsec$ from UV counterpart within the circle.  
\end{enumerate}
\begin{table*}
	\centering
	\caption{UV counterparts of X-ray sources.}
	
\begin{tabular}{|c|c|c|c|c|c|c|c|c}
\hline
\hline 
UV counterparts & \multicolumn{4}{c|}{NUV/N279N} & \multicolumn{4}{c|}{FUV/F169M} \\
 & Rate (cts sec$^{-1}$)$^a$ & flux$^b$ & flux$^c$ &$\dfrac{F_X^{d}}{F_{NUV}}$ &Rate (cts sec$^{-1}$)$^a$ &flux$^b$& flux$^c$ & $\dfrac{F_X^{d}}{F_{FUV}}$\\
\hline 
X2 & 0.28$\pm$0.004 & 0.98$\pm$0.02& 1.07$\pm$0.02&0.03 &0.47$\pm$0.004 & 2.06$\pm$0.02 & 2.25$\pm$0.02&0.02\\ 
X8 & 0.08$\pm$0.002 & 0.28$\pm$0.007 & 0.31$\pm$0.01& 0.02&0.14$\pm$0.002 & 0.61$\pm$0.009 &0.67$\pm$0.01&0.01\\
X13 & 0.32$\pm$0.004 & 1.12$\pm$0.02 & 1.22$\pm$0.02&0.01&0.52$\pm$0.005 & 2.28$\pm$0.02 &2.49$\pm$0.02&0.01\\
X17 & 0.06$\pm$0.002 & 0.21$\pm$0.007 &0.23$\pm$0.01&0.05&0.13$\pm$0.002 & 0.57$\pm$0.009 &0.62$\pm$0.01&0.03\\ 
\hline
\hline 
\end{tabular} 
    \\
\begin{flushleft}
    $^a$ Background corrected net count rates of UV counterparts extracted using a circular aperture of 3$\arcsec$ radius.\\
	$^b$ flux in units of 10$^{-15}$ ergs cm$^{-2}$ sec$^{-1}$ \AA$^{-1}$ \\
	$^c$ Galactic extinction corrected flux in units of 10$^{-15}$ ergs cm$^{-2}$ sec$^{-1}$ \AA$^{-1}$ \\
	$^d$ The Galactic absorption corrected F$_X$ value is taken from \cite{Strateva_2009}
\end{flushleft}
	\label{tab:tab25}
\end{table*}
\subsubsection{IR counterparts}
\begin{enumerate}[label=(\roman*)]
\item We found the high IR emission from X2 and X13 in the MIPS 24 and IRAC 8 $\mu$\rm{m} bands (see Table \ref{tab:tab25}). 
Additionally, the IR analogues of the hotspot regions are also present. 
The UV emission combined with IR emission shows star forming regions with PAH emission as well as hot dust emission. We found that X2 and X13 sources in IRAC 8 $\mu$\rm{m} band are two and three times wider than the FWHM of the IRAC instrument ($\sim$ 2$\arcsec$), respectively. However, X2 and X13 are extended sources in MIPS 24 $\mu$\rm{m} band. In summary, X2 which is a known ULX source has both UV and IR Counterparts and coincides with HII region. The detailed broadband study of these sources is beyond the scope of this paper.

\item Infrared emission is diffused and there is some modest UV emission from X8. As we can see, the X9 emits an extremely low magnitude of UV radiation, but there are also patches of infrared emission. 

\item When looking at X15, point source infrared emission is the only thing that stands out, with a FWHM of 2.79$\arcsec$. It indicates that it contains compact IR clusters with X-ray emission, which is identical to the FWHM of the IRAC instrument, which is $\sim$ 2$\arcsec$. It is likely to be a background AGN as already stated in \cite{Strateva_2009}.
\end{enumerate}

\subsection{The nuclear UV emission}
\label{sec3.3}
First, we performed aperture photometry of the central region of 6$\arcsec$ radius centered on NGC 1365 to check the variable nature of NGC 1365. For this, we calculated total counts from the source and counts from the average of five background regions of 6$\arcsec$ regions. Then the net counts of the source is calculated by subtracting the averaged background counts. The count rate for all the observations are listed in the first two columns of Table \ref{tab:tab3}. 

\begin{table}
	\centering
	\caption{UV/X-ray count rates of the nuclear region of NGC~1365.} 
	\centering
\begin{tabular}{|c|c|c|c|c|c|c|c|}
\hline
\hline 
OBS. ID & NUV$^a$& FUV$^b$ & SXT$^c$ \\
 & Rate (cts sec$^{-1}$)  & Rate (cts sec$^{-1}$)   & Rate (cts sec$^{-1}$)  \\
\hline 
9000000776 & 1.56$\pm$0.01 & 0.84$\pm$0.01  & 0.17$\pm$0.004\\ 
9000000802 & 1.43$\pm$0.01  & 0.74$\pm$0.008 & 0.27$\pm$0.004\\
9000000934 & 1.47$\pm$0.015 & 0.84$\pm$0.01 & 0.08$\pm$0.003\\
\hline
\hline 
\end{tabular}
	\\
	$^a$, $^b$ Background corrected net count rates and extinction$+$background corrected net count rates (Rate$_{corr}$) of NGC 1365 (AGN + host galaxy) extracted using a circular aperture of 6$\arcsec$ radius. \
	\newline
	$^c$ Background-corrected net SXT count rates in 0.3 $-$ 7 keV band.
	\label{tab:tab3}
\end{table}

\begin{figure*}
	\includegraphics[width=8cm, height=8cm]{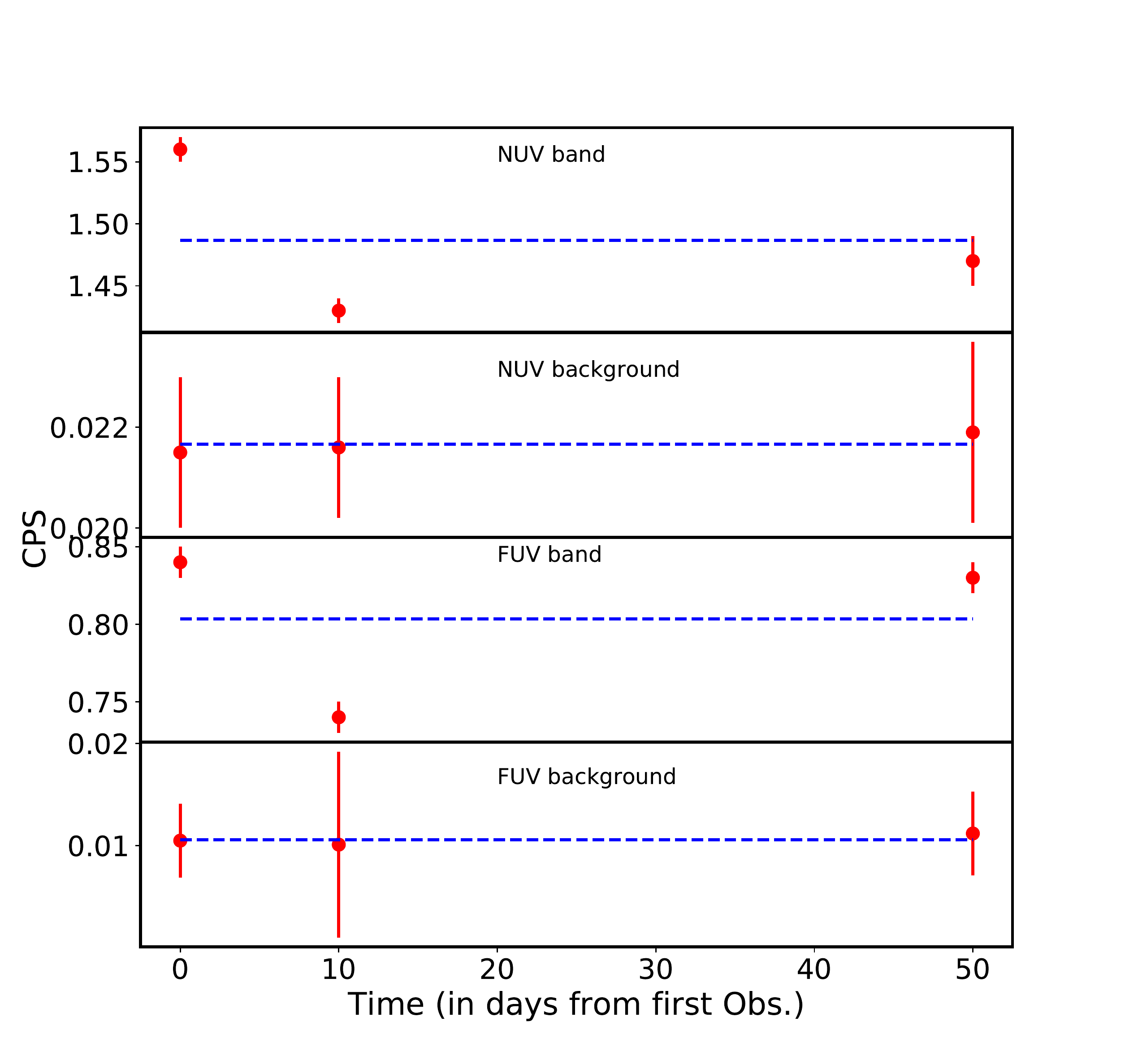}
	\includegraphics[width=8cm, height=8cm]{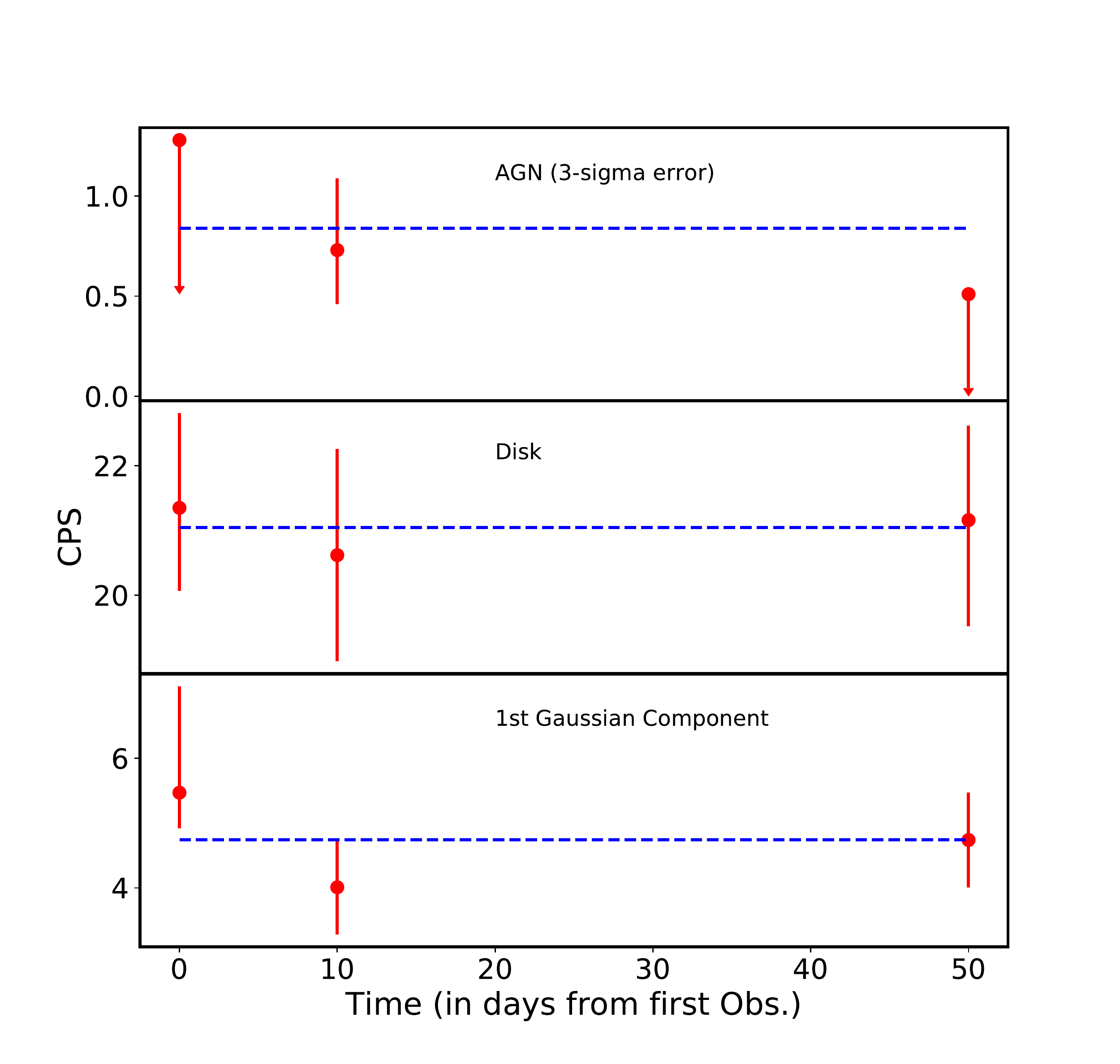}
    \caption{Left panel: The NUV and FUV light curves of NGC 1365 (AGN $+$ host galaxy) with the constant background CPS from 6$\arcsec$ aperture photometry (see Section \ref{sec3.3}). Right panel: The NUV light-curve of the AGN, disk, and Gaussian component using count rates derived
from the radial profile analyses (see, Table \ref{tab:tab9}). The Blue dashed line in each panel indicates the best fit constant count rate.}
    \label{fig:fig19}
\end{figure*}

The net source count rate and the background count rate in both NUV and FUV bands are plotted in Figure \ref{fig:fig19} (left panel). We checked the variability of the source by $\chi^{2}$
analysis, which measures the deviation of the data points
from the best-fit constant. The best-fit constants are shown as the horizontal blue dashed lines in
Fig. \ref{fig:fig19}. The NUV and FUV emission from the central regions of NGC 1365 appear to be variable. But it is not clear, whether the variability is caused by the AGN or not. However, we do not detect any variability from regions of the radius of 2" and 4". It is also possible that our extraction regions cover slightly different emission regions which may be possible due to slight error in the astrometry. The average accuracy of RA, Dec are $\sim$ 0.24", $\sim$ 0.55" and $\sim$ 0.53" and $\sim$ 0.48" in the NUV and FUV bands respectively. In general, the UV emission from an active galaxy includes the emission from the AGN and the host galaxy. Also, the observed flux is influenced by intrinsic and Galactic extinction. First, we employed the radial profile method to separate the host galaxy and AGN emission. Second, we corrected for the intrinsic and the Galactic extinction as described below.

  NGC 1365 is a spiral galaxy which is comprised of the active nucleus, disk, bulge, and bar. Each component can be separated by utilizing the excellent spatial resolution of the
UVIT. We construct and fit the radial profile of the galaxy to separate the host galaxy and AGN emission. We first determined the point spread function (PSF) of the instrument by fitting a Moffat function to the Radial profile of a point source. 
We extracted the radial profile of a field star and fitted a Moffat function $I_M$($1+((r-c)/\alpha)^2)^{-\beta}$. The radial profile is fitted using Sherpa 4.12.0 in Python and the errors calculated for each parameter correspond to the 90\% confidence range. The FWHM is calculated using the equation 2$\alpha$ $\sqrt{2^{1/\beta}-1}$.
The PSF in the NUV band is derived to be 1.34$\arcsec$, 1.57$\arcsec$ and 1.50$\arcsec$ for the 1st, 2nd \& 3rd observations, respectively. The PSF in the FUV band is derived to be 1.56$\arcsec$, 1.32$\arcsec$ and 1.49$\arcsec$, respectively, for the three observations. \\
\par
  Likewise, for each observation, we fitted the radial profile of
NGC 1365 with the corresponding PSF to account for
the AGN emission by using Moffat function. It is noted that the center of AGN is taken from {\it Chandra} position as the AGN, a point source, is not clearly detectable in the strong diffuse emission in the  UV bands. The disk is modeled using an exponential profile $I_d$($exp^{-r/r_d}$) where $I_d$ is the surface brightness of the disk at r=0$\arcsec$ and r$_d$ is the disk length, plus a Gaussian function $I_o$ ($exp^{{((r-c)/2fwhm})^2}$) for emission feature at distance larger than 4$\arcsec$ from the nucleus and a constant component for the background. The total model is given by 
\begin{dmath}
  I(r) = I_M(1 + ((r-c)/\alpha)^2)^{-\beta} + I_d(exp^{-r/r_d}) + I_o (exp^{{((r-c)/2fwhm})^2})+\rm{constant} 
\end{dmath}
where c is the position of the center peak.
We state that the well resolved components present in the
UVIT images of this galaxy (e.g., hot spots in central
region of the radius of 7$\arcsec$, the extended spiral arm etc.) can be studied, but this model is primarily used to separate the AGN flux.
 Consequently, to get acceptable values of
the fit statistics, we had to add 5\% systematic error to our
data, which resulted in increased error bars in the best fit parameters.
\par
\begin{figure*}
	\includegraphics[width=\columnwidth]{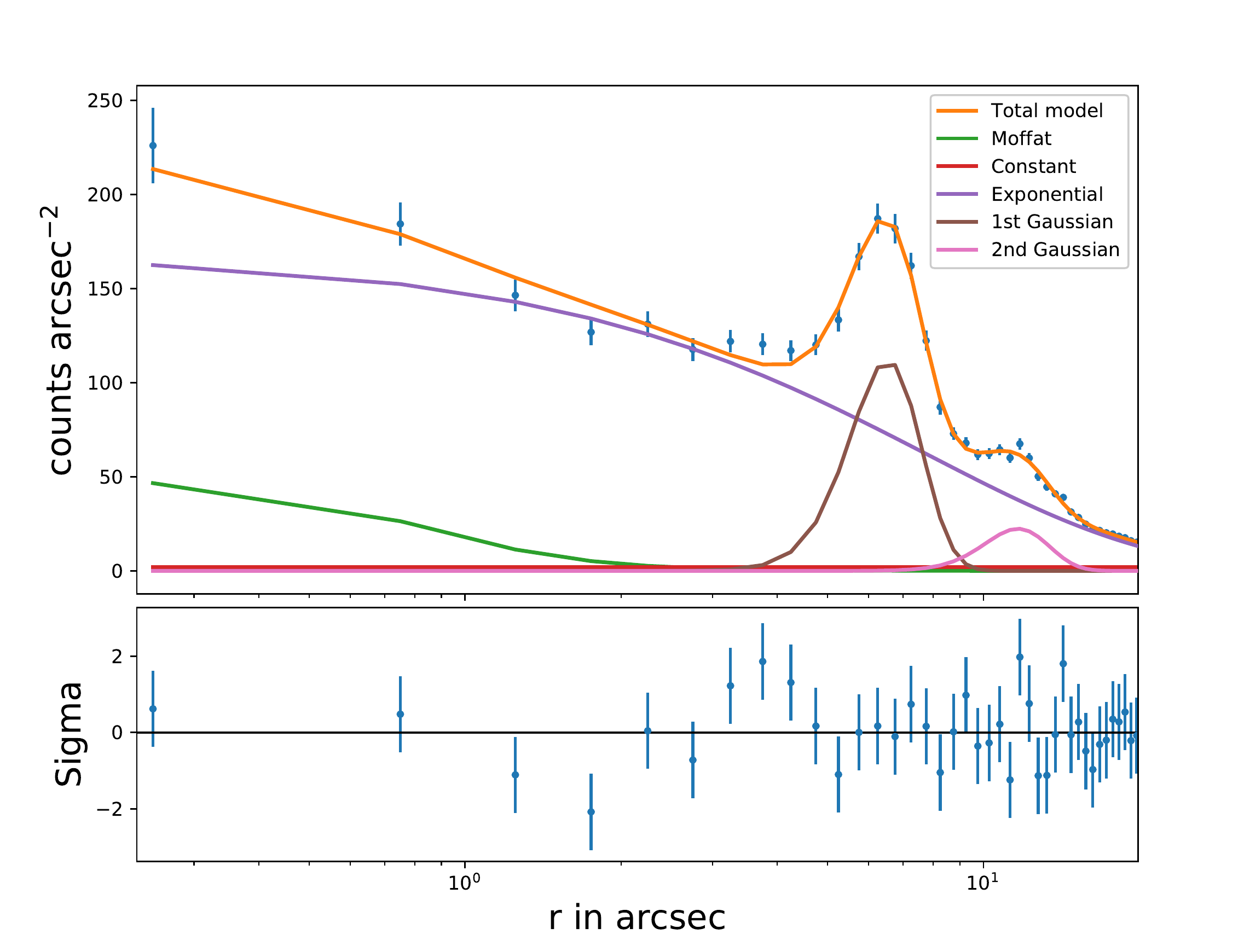}
	\includegraphics[width=\columnwidth]{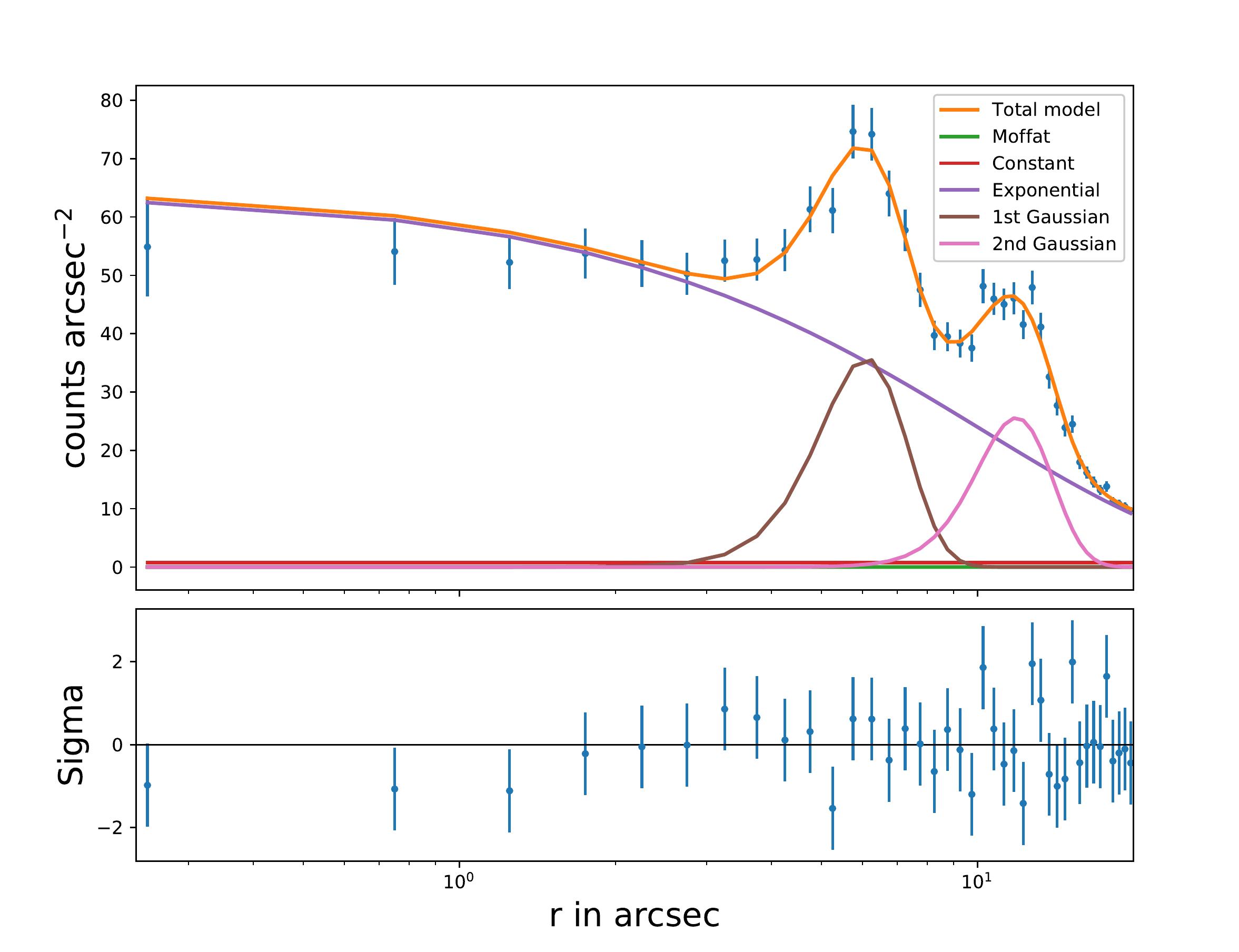}
    \caption{Radial profile of NGC 1365 in the NUV (left panel) and FUV (right panel) bands,  and the best fitting model consisting of a point source (PSF),  exponential function,  Gaussian function and a constant component.}
    \label{fig:fig15}
\end{figure*}

 The same procedure is also applied for the FUV band in order to get the AGN flux. 
 The radial profile for NGC 1365 for the first observation in both the bands are shown in Figure \ref{fig:fig15}. All the parameters are listed in Table \ref{tab:tab7} \& \ref{tab:tab8} with the best fit model parameters. The AGN is faint in the NUV band because of the heavy dust lane, but the nucleus in the FUV band is fainter than that in the NUV band. Hence, the host 
 galaxy contributions are larger than that in the NUV band.  
We integrated the
best-fit Moffat function, the exponential profile, and the
Gaussian component, over a radius of 6$\arcsec$,
and we calculated the  count rates for the AGN,
the disk, and the Gaussian components, respectively. The
count rates thus derived are listed in Table \ref{tab:tab9} (2nd, 3rd, 4th and 5th rows). 
\begin{table*}
	\centering
	\caption{Best-fit parameters from the NUV radial profile analysis for the source.}
	\label{tab:tab7}
\begin{tabular}{|c|c|c|c|c|c|c|c|c|}
\hline
\hline 
Component & Parameters &  Type &  & OBS\_ID & \\
& & & 9000000776 & 9000000802 & 9000000934 \\
\hline
Moffat &   A$_m$       &  thawed &    47.73$_{-29.98}^{+30.1}$ & 130.94$_{-39.5}^{+39.95}$ & $<$20.13\\
\\
  & c$_m$       &  freezed & 0.14 & 0 & 0\\
   \\
 &  s$_m$       &  freezed & 0.88 & 1.17 & 1.07\\
\\
  & b$_m$       &  freezed &   1.51 & 1.86 & 1.72\\
\hline
 Constant & co       &  freezed &    1.92  & 2.49 & 1.40\\ 
    \hline
Exponential  & ex$_{I}$      &  thawed &    168.61$_{-8.55}^{+8.46}$ & 192.93$_{-12.29}^{+12.06}$ & 109.38$_{-6.64}^{+5.78}$\\ 
\\  
  & rd$_{ex}$      &  thawed &    7.78$_{-0.20}^{+0.22}$ & 7.96$_{-0.27}^{+0.29}$ & 7.79$_{-0.24}^{+0.27}$\\ 
    \hline   
1st Gaussian  & A$_{G1}$       &  thawed &    112.07$_{-7.80}^{+7.95}$ & 120.21$_{-11.63}^{+12.05}$ & 69.38$_{-6.14}^{+6.28}$ \\ 
\\
  & c$_{G1}$     &  thawed &    6.52$_{-0.07}^{+0.07}$ & 6.72$_{-0.09}^{+0.09}$& 6.57$_{-0.09}^{+0.09}$\\
\\
  & fwhm$_{G1}$    &   thawed &   1.37$_{-0.10}^{+0.10}$ & 1.35$_{-0.14}^{+0.15}$ & 1.40$_{-0.13}^{+0.14}$\\ 
    \hline  
2nd Gaussian  & A$_{G2}$       &   thawed &   22.37$_{-2.25}^{+2.31}$ & 26.66$_{-3.22}^{+3.28}$ & 15.11$_{-1.81}^{+1.84}$\\ 
\\  
  & c$_{G2}$        &  thawed &    11.65$_{-0.21}^{+0.19}$  & 12.14$_{-0.26}^{+0.24}$ & 11.13$_{-0.43}^{+0.33}$\\
\\ 
  & fwhm$_{G2}$    &  thawed &    0.84$_{-0.09}^{+0.10}$  & 0.83$_{-0.10}^{+0.12}$ & 0.75$_{-0.12}^{+0.13}$\\
\hline 
  $\chi^2_{reduced}$/DOF & & & 1.01/31 & 0.99/31 & 1.08/31 \\
   \hline
   Systematic error & & & 0.05 & 0.055 & 0.05 \\
\hline
\hline
\end{tabular} 
\end{table*} 
   
\begin{table*}
	\centering
	\caption{Best-fit parameters from the FUV radial profile analysis for the source.}
	\label{tab:tab8}
\begin{tabular}{|c|c|c|c|c|c|c|c|c|}
\hline 
\hline
Component & Parameters &  Type &  & OBS\_ID & \\
 & & & 9000000776 & 9000000802 & 9000000934 \\
\hline
Moffat & A$_m$       &  thawed &    $<$8.05 & 22.64$_{-22.09}^{+22.22}$ & $<$8.25 \\
\\
 &  c$_m$       &  freezed & 0 & 0 & 0\\
\\
  & s$_m$       &  freezed & 0.86 & 0.88 & 0.84\\
\\
  & b$_m$       &  freezed &   1.14 &1.56 & 1.19\\
\hline
Constant &   co       &  freezed &    0.72  & 1.15 & 0.81\\ 
\hline
 Exponential & ex$_{I}$      &  thawed &    64.37$_{-7.19}^{+2.43}$ & 87.43$_{-6.38}^{+6.16}$ & 51.63$_{-8.43}^{+1.32}$\\ 
\\  
  & rd$_{ex}$      &  thawed &    10.51$_{-0.25}^{+0.87}$ & 10.85$_{-0.54}^{+0.62}$ & 10.09$_{-0.26}^{+1.32}$\\ 
\hline  
1st Gaussian & A$_{G1}$       &  thawed &    35.54$_{-3.32}^{+3.79}$ & 52.34$_{-5.32}^{+5.42}$ & 36.93$_{-3.57}^{+4.10}$\\ 
\\
  & c$_{G1}$       &  thawed &    6.08$_{-0.15}^{+0.14}$ & 6.54$_{-0.14}^{+0.14}$ & 6.02$_{-0.14}^{+0.12}$\\
\\
  & fwhm$_{G1}$    &   thawed &   1.18$_{-0.19}^{+0.13}$ & 1.09$_{-0.12}^{+0.13}$ & 1.28$_{-0.23}^{+0.12}$\\ 
    \hline
 2nd Gaussian & A$_{G2}$       &   thawed &   25.39$_{-1.55}^{+1.86}$ & 42.22$_{-2.93}^{+2.99}$ & 23.03$_{-1.47}^{+1.85}$\\ 
\\  
  & c$_{G2}$       &  thawed &    11.86$_{-0.17}^{+0.15}$ & 12.34$_{-0.15}^{+0.15}$ & 11.51$_{-0.23}^{+0.18}$\\
\\
  & fwhm$_{G2}$    &  thawed &    0.70$_{-0.05}^{+0.05}$  & 0.75$_{-0.06}^{+0.06}$ & 0.66$_{-0.04}^{+0.06}$\\
\hline
   $\chi^2_{reduced}$/DOF & & & 0.97/31 & 1.02/31 & 0.99/31 \\
   \hline
   Systematic error & & & 0.05 & 0.06 & 0.05 \\
 \hline
    \hline
\end{tabular} 
\end{table*}

The galactic disk is contributing the most in all the observations, implying that the central region is dominated by the host galaxy emission. However, overall the AGN in the NUV band is still detectable. The AGN emission in the FUV is undetectable, most likely hidden by the heavy dust lane.
We use the AGN count rates in the NUV and FUV bands further for extinction correction. 

\subsubsection{Extinction Correction}
We used the extinction curve of \cite{cardeli1989} for the Galactic extinction with A$_{V,Galactic}$ = 0.05 (NED) where A$_V$ is the extinction
in the V band. The AGN count rates are corrected for the
Galactic extinction.
Following \cite{davies2020ionized} with A$_V$ = 1.7, we used the
extinction curve of \cite{Calzetti_2000} for the intrinsic extinction. 
The actual value of dust extinction E(B-V) could be considerably 
more significant as we look through many clouds. 
The value of extinction is calculated by \cite{edmunds1982} as 
they have calculated color excess from NaI method by using 
various Doppler broadening values (b). We have used b value at 
0.3 as broadening parameter (b) $<$ 50 seems unlikely. This derived 
extinction value perfectly matches with
 \cite{davies2020ionized}'s extinction value with A$_V$ = 1.7.
Hence we have used this extinction value for our work. To get extinction in the respective filter, 
we used python package Extinction, which is available (\url{https://extinction.readthedocs.io/en/latest/api/extinction.calzetti00.html}) with R$_V$ = 4.05 for starburst galaxies. The AGN count rates corrected for Galactic \& intrinsic extinction are listed in the sixth row of Table \ref{tab:tab9}. We then
converted the intrinsic source count rates to flux density 
(in ergs cm$^{-2}$ sec$^{-1}$ \AA$^{-1}$ ), using the flux conversion factor
for the FUV/F169M and NUV/N279N filters \citep{Tandon2020}. The intrinsic fluxes of the source are given in the last row of Table \ref{tab:tab9}.
\begin{equation}
    A_{FUV}=R_{FUV}\times(E(B-V)).
	\label{eq:quadratic}
\end{equation}

\begin{figure}
    \centering
    \includegraphics[width=8cm,height=5cm]{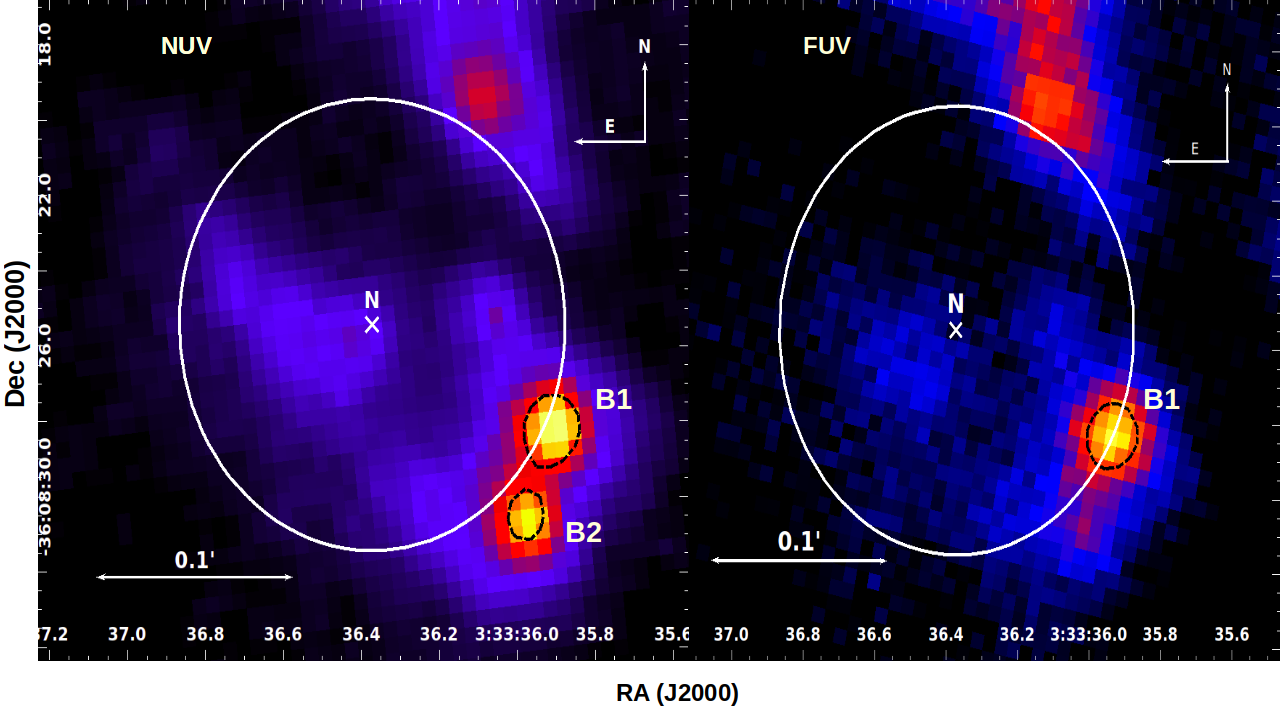}
    \caption{The UVIT NUV (left) and FUV (right)  image of NGC~1365  with circular centered region of radius 6" with bright spots B1 and B2.}
    \label{fig:fig51}
\end{figure}

\begin{table*}
	\centering
	\caption{Intrinsic count rates and flux of NGC 1365 in NUV and FUV band} 
	\label{tab:tab9}
\begin{tabular}{|c|c|c|c|c|c|c|c|c|}
\hline
\hline 
Component & &NUV & & &FUV & \\
 & 9000000776 &9000000802 &9000000934 & 9000000776 & 9000000802 &9000000934 \\
\hline
Background$^a$ & 0.02 & 0.02& 0.02& 0.01&0.01 &0.01  \\
\\
Disk$^a$ & 1.17$_{-0.07}^{+0.08}$ & 1.13$\pm$0.09 & 1.16$_{-0.09}^{+0.08}$ & 0.64$_{-0.08}^{+0.04}$ & 0.57$\pm$0.05 &0.59$_{-0.1}^{+0.04}$\\
\\
1st Gaussian$^a$ & 0.3$_{-0.03}^{+0.09}$ &0.2$\pm$0.04 &0.26$\pm$0.04 & 0.21$_{-0.02}^{+0.03}$& 0.15$\pm$0.03 & 0.24$_{-0.01}^{+0.03}$ \\
\\
2nd Gaussian$^a$($\times 10^{-4}$) &1.32$_{-1.15}^{+5.11}$ & 0.53$_{-5.06}^{+3.42}$& 12.3$_{-0.19}^{+56.2}$& 10.4$_{-6.3}^{+14.3}$& 2.21$_{-1.64}^{+4.97}$&31.1$_{-19.4}^{+30.8}$ \\
\\
AGN$^a$ & 0.03$\pm$0.02 & 0.04$\pm$0.01 & $<$ 0.01 & $<$0.009 & 0.01$\pm$0.009 & $<$0.01\\
\\
Galactic+Intrinsic  & 0.55$\pm$0.36& 0.73$\pm$0.18 & $<$0.18& $<$0.47& 0.47$\pm$0.46 & $<$0.52\\
extinction corrected$^a$ & & & & & & \\
\\
Intrinsic AGN flux$^b$  & 1.92$\pm$1.24 & 2.56$\pm$0.64 & $<$0.63 & $<$2.06 & 2.06$\pm$1.99 & $<$2.28 \\
\\

Disk flux$^b$ & 4.1$_{-0.25}^{+0.29}$& 3.96$\pm$0.32 &4.06$_{-0.33}^{+0.29}$ &2.81$_{-0.4}^{+0.18}$ & 2.5$\pm$0.22 & 2.59$_{-0.45}^{+0.18}$ \\
\\
1st Gaussian flux$^b$ & 1.05$_{-0.12}^{+0.32}$&0.71$\pm$0.14 &0.91$\pm$0.14 &0.92$_{-0.09}^{+0.13}$ &0.66$\pm$0.13&1.05$_{-0.04}^{+0.13}$\\
\\
2nd Gaussian flux$^b$ ($\times 10^{-4}$) & 4.62$_{-17.88}^{+4.03}$& 1.86$_{-17.76}^{+11.9}$ &43.05$_{-6.5}^{+196.72}$ &45.66$_{-27.4}^{+62.77}$ &9.70$_{-7.19}^{+21.82}$ &136.53$_{-85.26}^{+135.16}$ \\ 
\hline
\hline
\end{tabular} 
\\
\begin{flushleft}
$^a$ In cts sec$^{-1}$  \\
$^b$ In units of 10$^{-15}$ ergs cm$^{-2}$ sec$^{-1}$ \AA$^{-1}$  \\
\end{flushleft}
\end{table*}
Fig. \ref{fig:fig19} (right panel) shows the AGN, disk and 
Gaussian component light curves in the
NUV/N279N band (from top to bottom, respectively),
using the values listed in Table \ref{tab:tab7}.  
Clearly, the NUV count rate is not variable within 3-$\sigma$ error.
The emission from the disk is almost constant over the observations period whereas
the emission from Gaussian component is slightly varying. Thus, the 
emission from the Gaussian components 
 is consistent with the results plotted in the left panel of \ref{fig:fig19} (1st plot).
 It shows that the variation in UV count rate from 6$\arcsec$ aperture are most likely due to the variations in the Gaussian components. The AGN in the FUV band is not well detected and remains constant within the upper-limits, hence we have not plotted. 
 Since we employed a nuclear region with a 6$\arcsec$ aperture size, the errors in the astrometric corrections of bright spots at the edge of the extraction region as shown in Figure \ref{fig:fig51} might cause the variation of UV flux. We have not examined the details of these two bright spots in this study.\\

\subsection{SXT spectral analysis}
\label{sec3.4}
\begin{figure}
	\includegraphics[width=5.5cm,angle=-90]{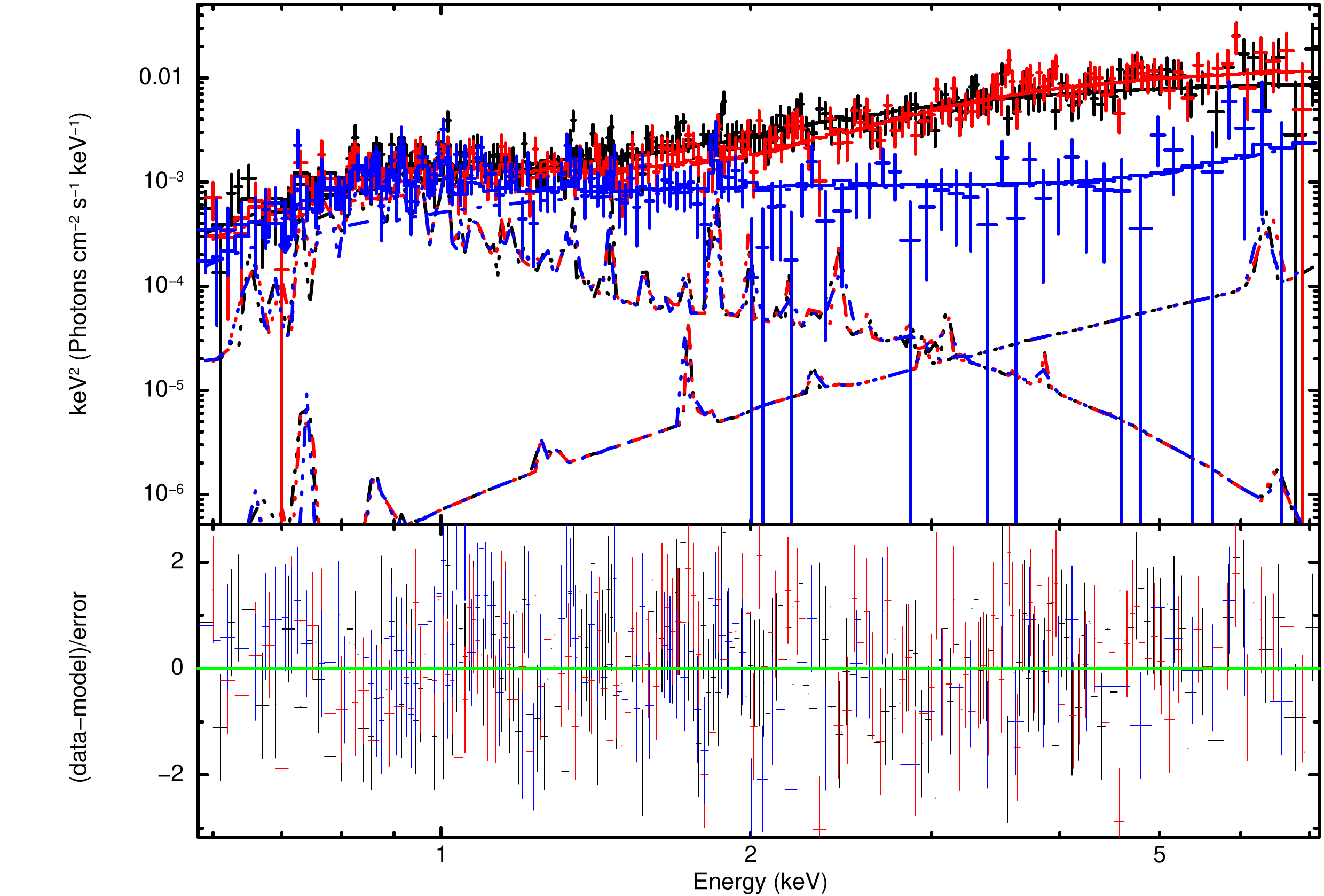}
    \caption{The SXT spectral data from the three observations (distinguished with different colors: black, red, blue in chronological order), together with the best-fitting model ({\sc tbabs({xillver} $+$ apec $+$ ztbabs $\times$ tbpcf $\times$ powerlaw}) (solid lines of the corresponding colors) and all the components (dashed lines of the corresponding colors). The residuals are reported in the bottom panel.}
    \label{fig:fig17}
\end{figure}

We fitted the three sets of SXT spectral data (see Table~\ref{tab:tab1}) in the 0.6 $-$ 7 keV band using XSPEC version 12.8.2. 
We employed the $\chi^2$ statistics and calculated errors on the best fit parameters at 90\% confidence level, unless otherwise stated.
We note that due to slight change in SXT gain \citep{2017JApA...38...29S}, we shifted the energies in the SXT RMF and ARF using the $XSPEC$ gain fit. We fixed gain slope to 1 and varied the offset parameter. Then we fixed the offset parameter at 0.03.
Also, we applied 3\% systematic error in the spectral models to account for calibration uncertainties. \\
First we fitted the three sets of
SXT observations jointly in 2 $-$ 7 keV band of the spectrum with a {\sc powerlaw} where the photon index and normalization of powerlaw were left free to vary. The Galactic column density is fixed to $N_{\rm H}$ = 1.34$\times$10$^{20}$ cm$^{-2}$ \citep{kalberla2005}. The photon indices are 0.86$^{+0.17}_{-0.17}$,
0.39$^{+0.17}_{-0.17}$ and 0.97$^{+0.78}_{-0.78}$ respectively and normalization (norm$_{PL}$) are 0.001$^{+0.0003}_{-0.0002}$, 0.0007$^{+0.0002}_{-0.0001}$ and 0.0003$^{+0.0003}_{-0.0002}$ in units of photons keV$^{-1}$ cm$^{-2}$ sec$^{-1}$ at 1 keV respectively with reduced $\chi^2$ ($\chi^2_{red}$) 0.96 for 200 degree of freedom (DOF) with null probability 0.7. We extrapolate the data upto 0.7 keV, which resulted in a poor fit with residuals mostly in the soft band. In order to improve the fit, blackbody component is added to the model, but still gives a poor fit.
Instead of blackbody component, we used 
{\sc apec} model of AtomDB version 10 \citep{Smith_2001} to fit the soft excess for the thermal diffuse emission from the central regions. Following the approach of \cite{Wang_2009},
the abundances parameter is kept fixed at 1.
 After including {\sc apec} model, the $\chi^2_{red}$ is improved to 0.91 for 461 DOF with null probability 0.92. We fixed the plasma temperature at 0.8 kT and kept tied the normalization of {\sc apec} component. We obtained a similar fit as in the previous case
from a statistical point of view (overall $\chi^2_{red}$ 0.91). Now the photon index of third observation is improved to 1.91$^{+0.23}_{-0.23}$. But the first two spectrum 
still remain flat which is unphysical. We multiplied an intrinsic neutral absorption model {\sc ztbabs} at the source redshift to the {\sc powerlaw} component,
and we kept the absorption column tied among all three
datasets. 
We then used the reflection model {\sc xillver} \citep{garcia2014} to account for related X-ray reflection emission. We tied the photon index of the reflection component with those of the power-law 
component for each observation, and we kept the normalization tied across 3 datasets. We
fixed the iron abundance at A$_{Fe}$ = 1, inclination angle
at $\theta$ = 60$^{\circ}$ , ionization parameter at log $\varepsilon$ = 0, cut-off
energy at E$_{cut}$ = 100 keV and
reflection fraction at R$_f$ = $-$1. 
NGC 1365 is known to show partial covering X-ray absorption \citep{Risaltti2009}, \citep{maiolino2010}, therefore we multiplied with a partial covering model {\sc tbpcf} to the powerlaw component with varying N$_H$ and f$_c$. We first consider a uniform absorbing cloud before varying the parameters in the {\sc tbpcf} model. We fixed the covering fraction at 1 and varied the column density. We obtained a poor fit ($\chi^2$/dof = 509.5/466) with column densities ranges in the order of $\sim$ 10$^{21}$ cm$^{-2}$ with $\Gamma$ at 0.89$_{-0.09}^{+0.21}$. Then, we varied the covering fraction and column density in the next step. We obtained a good fit with $\chi^2_{red}$ = 0.95 for 463 DOF with null probability 0.7. However, the partial-covering model is the most likely scenario from a physical point of view. We found the variation in N$_H$ value of 10$^{22-23}$ cm$^{-2}$ with $\Gamma$ at 2.25$_{-0.40}^{+0.35}$. We now fixed the covering fraction for the third observation to 0.8 as it was not well constrained. 
 We obtained the same $\chi^2_{red}$ = 0.95 for 464 DOF. The best fit parameters together with the associated errors (90\% confidence intervals) are listed in Table \ref{tab:tab4}.
 The spectra with best fit models are shown in Figure \ref{fig:fig17}, where the bottom panel shows the residual in terms of sigmas. The count rates, fluxes and luminosity are given in Table \ref{tab:tabcps}. The flux in second observation is 1.5 times and 3 times more than second and third observations.

\begin{table*}
	\centering
	\caption{Model parameters in {\sc tbabs({xillver} $+$ apec $+$ ztbabs $\times$ tbpcf $\times$ powerlaw}) model for SXT observations.}
	\label{tab:tab4}
	\begin{tabular}{lllllllllll} 
	\hline
	\hline
Obs. &  {\sc N$_{H, ztbabs}^a$} &  {\sc Norm$_{Xillver}^b$} & kT(keV) & {\sc Norm$_{Apec}^c$} & {\sc N$_{H,\sc tbpcf}^a$} & f$_{cover}$ & Photon index($\Gamma$) &  {\sc Norm$_{powerlaw}^d$} & log(f$_{pl}$) & $\chi^2$/DOF, p \\
 & & $\times$(10$^{-5}$) & & $\times$(10$^{-4}$) &&&& \\
		\hline
		\hline
\multicolumn{10}{c}{Model A}\\
\hline
1st & 0.34$_{-0.20}^{+0.27}$ & 6.36$_{-5.98}^{+6.72}$ & 0.8(f) & 2.15$_{-0.79}^{+0.77}$ & 5.44$_{-1.94}^{+2.97}$&  0.81$_{-0.08}^{+0.05}$ & 2.25$_{-0.40}^{+0.35}$ & 0.01$_{-0.006}^{+0.01}$ 
& -- 10.71$_{-0.08}^{+0.12}$ & 447.12/464, 0.71 \\
\\
2nd &  0.34(t) &  6.36(t) & 0.8(f) & 2.15(t) & 8.73$_{-2.14}^{+2.27}$ & 0.89$_{-0.05}^{+0.03}$ &  2.25(t) & 0.02$_{-0.01}^{+0.02}$ & -- 10.55$_{-0.09}^{+0.1}$ \\
\\
3rd &  0.34(t) &  6.36(t) & 0.8(f) & 2.15(t) & 73.98$_{-33.81}$ & 0.8(f) &  2.25(t) & 0.006$_{-0.002}^{+0.003}$ &-- 11.07$_{-0.07}^{+0.12}$ \\
\hline
\hline
\end{tabular}
\\ 
\begin{flushleft}
    $^a$ In units of 10$^{22}$ cm$^{-2}$. \\
	 $^b$ Normalization of the {\sc xillver} model.\\
     $^c$ Normalization of the {\sc Apec} model in units  of {10$^{-14}$ / 4$\pi$ [D$_A$ (1 +z )]$^2$}$\int n_e n_H \,dV$, where D$_A$ is the angular size distance to the source (cm), dV is the volume (cm$^3$) and n$_e$ and n$_H$ are the electron and H densities (cm$^{-3}$)\\
     
 $^d$ Power-law normalization in units of photons keV$^{-1}$ cm$^{-2}$ sec$^{-1}$ at 1 keV.\\   
\end{flushleft}
\end{table*}

\begin{table}
	\centering
	\caption{X-ray details of NGC~1365.} 
\begin{tabular}{|c|c|c|c|c|c|}
\hline
\hline 
OBS. ID & Count rate$^a$ & flux$^b$ & flux$^c$ & flux$^d$& luminosity$^e$ \\
        & (cts sec$^{-1}$) & & & &  ($\times$10$^{42}$ergs sec$^{-1}$)\\
\hline 
1 & 0.17$\pm$0.004 & 1.59 & 1.77 & 1.95& 0.75 \\ 
2 & 0.27$\pm$0.004 & 1.64 & 1.99& 2.82& 1.09 \\
3 & 0.08$\pm$0.003 & 0.41 & 0.34 & 0.86 & 0.33 \\
\hline
\hline 
\end{tabular}
	\newline
 \begin{flushleft}
 	The fluxes are in units of $\times$10$^{-11}$ergs sec$^{-1}$ cm$^{-2}$).\\
	$^a$ Background-corrected net SXT count rates in 2 $-$ 7 keV band.\\
	$^b$ Absorbed X-ray flux in 0.6 $-$ 7 keV band. \\
	$^c$ Absorbed X-ray flux in 2 $-$ 7 keV band.\\
	$^d$ Intrinsic X-ray flux in 2 $-$ 7 keV band.\\
	$^e$ Intrinsic X-ray luminosity in 2 $-$ 7 keV band.\\
 \end{flushleft}
	\label{tab:tabcps}
\end{table}

\section{Discussion}
We examined three simultaneous UV/X-ray observations
of NGC~1365 performed with \textit{AstroSat} during November -- December 2016. First, we analyzed the UV emission from the nuclear as well as the circumnuclear region.
We found two bright hot spots at 7$\arcsec$ from the
nucleus in the NUV band and  one bright spot in the FUV band. 
They are clearly associated 
with IR emitting regions and H$\alpha$ region. It suggests that these regions are 
ionized by UV radiation. These are high star forming regions with less AGN contribution \citep{agostino2018}. 
We identified four UV counterparts  and five IR counterparts of off-nuclear X-ray sources detected with {\it Chandra}. Most of the X-ray sources are collocated with HII extended regions. One of them, NGC 1365 X2, a well known ULX source from Fig. \ref{fig:fig10} has a bright UV spot within the circle of radius 3$\arcsec$ (270 pc) suggesting that ULX may be responsible for the observed UV emission. It is possible that the host environment of ULX X2 is a globular cluster with more than one stellar population, rather than an older globular cluster. Rest of the X-ray sources are coinciding with UV/IR extended emission suggesting association with star forming regions. More detailed study is required to firmly establish the association. 

Secondly, the excellent spatial resolution
of UVIT allowed us to separate the
AGN from the host galaxy emission by fitting the radial profile of NGC 1365 for each observation separately in the NUV and FUV bands.
We found  no substantial variation in the UV emission from the AGN  at a 2$\sigma$ level in the NUV band. Thus, the intrinsic NUV flux of NGC 1365 is not significantly variable  over the two-month period. The possible astrometry error of nearly 1 pixel (= 0.4$\arcsec$) of bright spots detected in UVIT observation, might be responsible for apparent variation of the UV emission from the central regions within  6$\arcsec$.
The star formation is more active along the spiral arms whereas it has been
suppressed in the central regions. This suppression could be linked to AGN feedback, which we plan to investigate in a future paper. 

Third, we found that the X-ray emission of NGC~1365 consists of a powerlaw component  with a photon index of 2.25$_{-0.40}^{+0.35}$ that is modified by partial covering  with variable column density N$_H$. Our data are consistent with the presence of diffuse emission in soft X-ray band with temperature kT $\sim$ 0.8 keV and X-ray reflection from a distant material. The X-ray spectral shape of NGC~1365 inferred is generally consistent those reported earlier \citep{Risaltti2009, maiolino2010}. We find strong variations in the column density of the partial covering absorbers. The source was Compton thin in the first two absorption with $N_H \sim 5-10\times10^{22}{\rm~cm^{-2}}$, while the absorption column for third observation increased at least by a factor four ($N_H > 4\times10^{23}{\rm~cm^{-2}}$), making the source nearly Compton-thick. NGC~1365 is known to show rapid transition between the Compton-thin and Compton-thick states during 6 weeks \citep{Risaliti_2005}. We also found the same as expected. We note that while X-ray absorption column varied strongly, we did not observe significant variability in the NUV emission. This suggests that the absorber is  covering the compact X-ray source but not the NUV emission arising from the disk. This is possible if the absorbing clouds themselves are compact and/or located not too far from the central source, but do not obscure the extended region. As suggested by \cite{maiolino2010}, the absorbing clouds may be located within the broad-line region. 
Furthermore, we applied the complex model {\sc tbabs({powerlaw$_{uncov}$} $+$ mekal $+$ gauss $+$ zphabs $\times$ zpcfab $\times$ powerlaw}) used by \cite{braito_2013} on SXT observations. SXT data requires a full covering layer of neutral absorbing material with a low column density of
1.36$_{-0.36}^{+0.38}$ $\times$ 10$^{22}$ cm$^{-2}$ and an emission peak at 1.07$_{-0.04}^{+0.03}$ keV with width $<$ 0.07 eV. It might be due to presence of the oxygen+iron+neon (O+Fe+Ne) emission blend possibly from diffuse emission\citep{Wang_2009}. We found the best fit photon index ($\Gamma$) at 1.91$_{-0.38}^{+0.46}$ which is within the range of $\Gamma$ obtained from \cite{braito_2013} and the present model used in this work. The spectral variability is mainly caused by varying X-ray column density (N$_H$ $\sim$ 10$^{22}$ - 10$^{23}$ \rm{cm}$^{-2}$) of obscuring absorber with covering fraction (f$_{cover}$) 0.64$_{-0.15}^{+0.12}$, 0.8$_{-0.07}^{+0.05}$ and 0.8 (fixed) for three observations respectively, which are close to the best fit value of f$_{cover}$ found by \cite{braito_2013}. Furthermore, the best fit values of column densities and covering fraction reported in this work are in agreement with \cite{Walton_2014} where $>$ $\sim$ 80\% of the clouds cover the X-ray source. The scenario of the partial covering model with self consistent disk reflection model suggested by \textit{NuSTAR} data \citep{risaliti2013rapidly, Walton_2014} supports our results of less than 7 KeV with no change in the flux from the reflection model.
\subsection{Determination of spectral energy distribution}
We have used simultaneous data on NGC~1365 in the X-ray and UV bands from \textit{AstroSat} and archival IR band data. Using the ftflx2xsp tool, we converted the intrinsic UV flux from Table 6 and IR flux into \rm{XSPEC} compatible spectral files for each observation. The AGN SED is constructed for a single observation (obsID: 776). We did not use the NIR data as this band is more likely to be affected by star formation in the central regions.
\begin{figure}
	\includegraphics[width=\columnwidth]{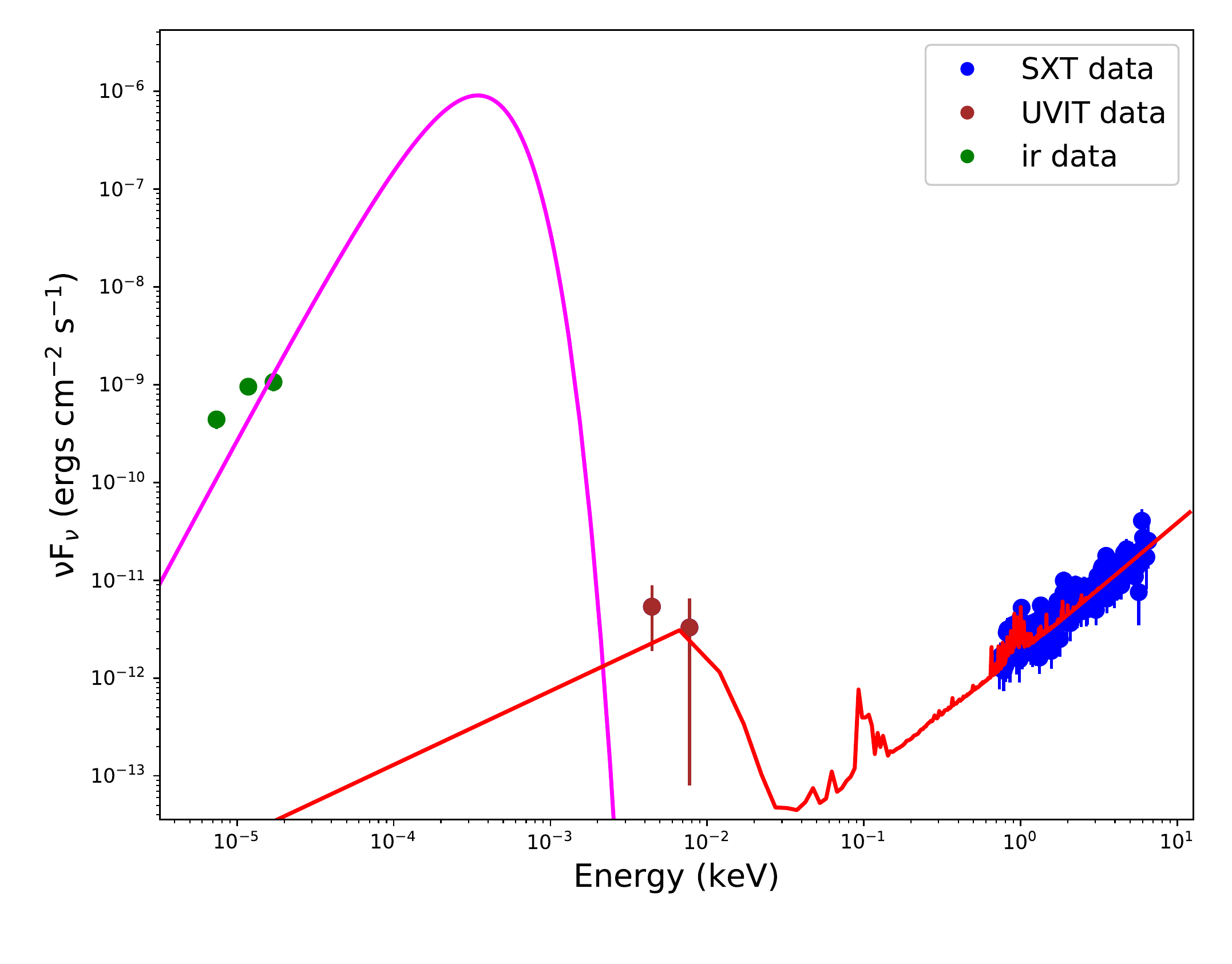}
    \caption{The intrinsic X-ray/FUV/NUV/IR SEDs for one of the observation (obsID 776) fitted with model {\sc constant (apec $+$ powerlaw $+$ disko)} (Red line) and blackbody model (Magenta line).}
    \label{fig:fig28}
\end{figure}
We fitted the X-ray data of 0.6 -- 7 keV band with {\sc apec} model for soft band in addition to 
the {\sc powerlaw} model for hard band. We have excluded Galactic absorption and absorption 
column density along the LOS inorder to get the primary X-ray continuum which is flat. Again we added a {\sc disko} 
model, which provides a multi-temperature of the blackbody accretion disk. We fixed the inner disc radius at r$_{in}$ = 3r$_g$ (r$_g$ = GM/c$^2$) and the standard viscosity ($\alpha$) 
 parameter at 0.01 \citep{king2007} and kept the accretion rate and its normalization n$_{DISKO}$ as variable parameters. The {\sc disko} normalization is defined as n$_{\sc disko}$ = 2cos$\theta$/D$^2$ where `$\theta$' is the inclination of the disk and D is the distance of the source. We adopted SMBH
mass $\sim$ 2 $\times$ 10$^6$ M$_{\odot}$ from \cite{risaliti2013rapidly}. 
Finally, we employed the physical model {\sc constant (apec $+$ powerlaw $+$ disko)}. We found a good $\chi^2$ value 154.39/180 DOF with null probability 0.92. The IR data are fitted with a blackbody component. The best fit AGN SED is shown in Figure \ref{fig:fig28}. The main parameters of the {\sc disko} model are the accretion rate and normalization (n$_{\sc disko}$) parameters where the accretion rate is constrained to 0.01$_{-0.003}^{+0.002}$ L$_{Edd}$ and n$_{\sc disko}$ at 1.61$_{-0.37}^{+0.51}$ $\times$ 10$^{-7}$. We obtained photon index 0.65$_{-0.09}^{+0.09}$ with its normalization 0.001$_{-0.0001}^{+0.0001}$ in units of 1/${D_{10}^2}$ where D$_{10}$ is the distance to the source in units of 10 kpc and {\sc apec} temperature 1.02$_{-0.17}^{+0.27}$ keV with its normalization at 2.01$_{-0.89}^{+0.89}$ $\times$ 10$^{-4}$. 
The SED also suggests that the AGN with low accretion rate $\sim$ 0.01 L$_{Edd}$ is less radiative, which means that UV and IR emission should likewise be low. However, the high IR in the AGN SED with weak accretion disk indicates emission from the circumnuclear ring surrounding the nucleus and from the torus around the accretion disk. 
\section{Conclusion}
We have performed a multi-wavelength study on the nearby barred
spiral galaxy NGC~1365, using high resolution far and near UV images available with the  UVIT, in conjunction with the spectral
capabilities of SXT, and studied the X-ray/UV properties of
the nuclear emission. We have also
analyzed the off-nuclear Chandra X-ray point source  with UV/IR counterparts in the galaxy. Our main results can be  summarized as
follows.\\
We separated the AGN emission and the host galaxy emission in the NUV band but did not detect the AGN in the FUV band. We also detected two bright spots at a distance of $\sim$ 7$\arcsec$ from the
nucleus in the NUV band, which is between the inner ILR ($\sim$ 3$\arcsec$) and outer ILR ($\sim$ 30$\arcsec$) in NGC 1365 \citep{1999A&ARv...9..221L}. 
We also found four UV counterparts and five IR counterparts of off-nuclear bright  X-ray sources detected with {\it Chandra}.
NGC 1365 X2 is a well-known ULX source featuring a UV bright spot and extended IR emission. This 
dense environment may be required for an ULX to have IMBH and may be hosted by young globular clusters. 
The other three sources X8, X13 and X17 are also coinciding with an extended emission in UV and IR.
The variation in the UV emission from the central regions could be caused by the error of $\sim$ 1 pixel from astrometry correction. We observed strong variability in X-rays that is mainly caused by varying column density. At the same time, the NUV emission from the disk are not covered by the absorbing clouds suggesting the compactness of clouds situated near to the central region. The strong IR emission and relatively weaker UV/X-ray emission as seen in the SED suggest that the central region contributes substantial IR emission likely arising from star formation.

\section*{Acknowledgements}

This publication uses
the data from the AstroSat mission of the Indian Space
Research Organisation (ISRO), archived at the Indian
Space Science Data Centre (ISSDC). This publication
uses the data from the UVIT and SXT.  The
SXT was processed by the pipeline
software provided by the respective payload operation
centers (POCs) at TIFR, Mumbai. The UVIT data
were checked and verified by the POC at IIA, Bangalore, and 
processed by the CCDLAB pipeline \citep{Postma_2017}. The archival IR data 
are used from $Spitzer$ and \textit{Herschel} telescope. This research has made use of the python
packages. This research has made use of the NED database. One of the authors (PS) acknowledges Manipal Centre for Natural Sciences, Centre of Excellence, Manipal Academy of Higher Education (MAHE) for facilities and support.

\section{Data Availability}
The Level-1 UVIT and SXT data from the three observations with
IDs 9000000776, 9000000802 and
9000000934 used in this paper are publicly available at the
AstroSat data archive \url{https://astrobrowse.issdc.gov.in/astro_archive/archive/Home.jsp} maintained by the ISSDC. 



\bibliographystyle{mnras}
\bibliography{reference} 



\appendix

\section{Some extra material}

If you want to present additional material which would interrupt the flow of the main paper,
it can be placed in an Appendix which appears after the list of references.


\bsp	
\label{lastpage}
\end{document}